\documentclass[fleqn,usenatbib]{mnras}

\usepackage[T1]{fontenc}

\DeclareRobustCommand{\VAN}[3]{#2}
\let\VANthebibliography\thebibliography
\def\thebibliography{\DeclareRobustCommand{\VAN}[3]{##3}\VANthebibliography}

\usepackage{graphicx}	
\usepackage{amsmath}	
\usepackage{amssymb}	

\usepackage{newtxtext,newtxmath}

\usepackage[utf8]{inputenc} 
\usepackage{listings}
\usepackage{hyperref}
\usepackage{setspace} 
\usepackage{siunitx} 
\usepackage{booktabs}
\usepackage{placeins}
\usepackage{threeparttable} 
\usepackage{tikz,tikz-3dplot} 
\usetikzlibrary{positioning,math,calc}

\newcommand{\Rey}{{\rm Re}}
\newcommand{\Pec}{{\rm Pe}}
\newcommand{\Ric}{{\rm Ri}}
\newcommand{\Ray}{{\rm Ra}}
\newcommand{\Pra}{{\rm Pr}}
\newcommand{\Ros}{{\rm Ro}}
\newcommand{\alignedb}{\begin{aligned}}
\newcommand{\alignede}{\end{aligned}}
\newcommand{\unitvect}{\hat {\mathbf e}}

\title[Modeling coexisting GSF and shear instabilities in rotating stars]{Modeling coexisting GSF and shear instabilities in rotating stars}

\author[E. Chang \& P. Garaud]{
Eonho Chang,$^{1,2}$\thanks{E-mail: eonhochang@math.arizona.edu; \quad pgaraud@ucsc.edu}
Pascale Garaud,$^{1}$\footnotemark[1]
\\
$^1$Department of Applied Mathematics, Baskin School of Engineering,
University of California, Santa Cruz, CA 95064 \\
$^2$Graduate Interdisciplinary Program in Applied Mathematics,
University of Arizona, Tucson, AZ 85721 
}

\date{Accepted XXX. Received YYY; in original form ZZZ}

\pubyear{2021}

\begin{document}
\label{firstpage}
\pagerange{\pageref{firstpage}--\pageref{lastpage}}
\maketitle

\begin{abstract}
     Zahn's widely-used model for turbulent mixing induced by rotational shear has recently been validated (with some caveats) in non-rotating shear flows. It is not clear, however, whether his model remains valid in the presence of rotation, even though this was its original purpose. Furthermore, new instabilities arise in rotating fluids, such as the Goldreich-Schubert-Fricke (GSF) instability. Which instability dominates when more than one can be excited, and how they influence each other, were open questions that this paper answers. To do so, we use direct numerical simulations of diffusive stratified shear flows in a rotating triply-periodic Cartesian domain located at the equator of a star. We find that either the GSF instability or the shear instability tends to take over the other in controlling the system, suggesting that stellar evolution models only need to have a mixing prescription for each individual instability, together with a criterion to determine which one dominates. However, we also find that it is not always easy to predict which instability ``wins'' for given input parameters, because the diffusive shear instability is subcritical, and only takes place if there is a finite-amplitude turbulence ``primer'' to seed it. Interestingly, we find that the GSF instability can in some cases play the role of this primer, thereby providing a pathway to excite the subcritical shear instability. This can also drive relaxation oscillations, that may be observable. We conclude by proposing a new model for mixing in the equatorial regions of stellar radiative zones due to differential rotation.
\end{abstract}

\begin{keywords}
hydrodynamics --- instabilities --- turbulence --- stars:evolution
\end{keywords}



\section{Introduction}
\label{sec:intro}

The theory of stellar evolution models the life of stars from their birth to their death, and is generally successful in reproducing most salient aspects of the Hertzsprung-Russel (HR) diagram. Discrepancies between model and observations, however, usually emerge when probing more specific aspects of stellar evolution using photospheric chemical abundances \citep[see e.g.][]{Pinsonneault_ARAA_1997} or asteroseismology \citep[see e.g.][]{Aerts_astero_2021}, and are often resolved by invoking some amount of extra mixing in the star's radiative zone.
There are many possible sources of extra mixing in stars, as discussed for instance by \cite{zahn_rotational_1974}. Of particular interest in recent years are the diffusive (alternatively called secular) shear instability \citep{zahn_rotational_1974,zahn_circulation_1992,prat_turbulent_2013,Prat_mixing_2014,garaud_stability_2015,garaud_turbulent_2016,prat_mixing_2016,garaud_turbulent_2017,gagnier_turbulent_2018}, and the Goldreich-Schubert-Fricke (GSF) instability \citep{goldreich_differential_1967,fricke_instabilitat_1968,Knobloch_stability_1982,knobloch_nonlinear_1982,Korycansky_gsf_1991,Rashid_gsf_2008,barker_angular_2019,barker_angular_2020}, that can extract energy from the differential rotation of the star to drive turbulence, and therefore transport of chemical species and angular momentum. In what follows, we begin by briefly reviewing what is known about both types of instabilities and their transport properties, and then discuss what outstanding issues remain to be studied. We ignore magnetic fields, for simplicity, but acknowledge that they would in practice form an important part of the complete story. Again for simplicity, we consider only the equatorial region of a star, noting that the non-equatorial case is substantially more complicated  \citep{Knobloch_stability_1982,barker_angular_2020}. Near the equator, by symmetry, the angular velocity $\Omega$ is a function of the radius $r$ only, so that we can assume that $\Omega = \Omega(r)$. 

\subsection{Shear instabilities}

It has long been known \citep{richardson_supply_1920,miles_stability_1961,howard_note_1961} that non-diffusive (adiabatic) shear instabilities only grow when the local Richardson number $J$ drops below a constant of order unity somewhere in the flow, where
\begin{equation}
    J=\frac{N^2}{S^2},
    \label{eqn:ric}
\end{equation}
$N$ is the Brunt-V\"ais\"al\"a frequency and $S$ is the local shear (which would be equal to $S = S_\Omega = rd\Omega/dr$ for rotational shear). 
The criterion has a simple energetic interpretation: for instability to occur, the kinetic energy extracted by the perturbations from the mean flow, which is proportional to $S^2$, must exceed the potential energy lost in mixing the stratified fluid, which is proportional to $N^2$. In practice, however, this criterion is rarely satisfied \citep[see][for a simple explanation]{garaud_journey_2021}, except very close to the edge of a convective region where $N \rightarrow 0$. As such, standard adiabatic shear instabilities are almost never excited in radiation zones.

Crucially, \cite{townsend_effects_1958} noted that the Richardson criterion is relaxed by non-adiabatic effects, because the stabilizing role of thermal stratification is reduced by the fluid parcel exchanging heat with its surroundings.  
Adapting Townsend's results, \cite{zahn_rotational_1974} proposed a modified criterion for diffusive shear instabilities applicable to optically thick stellar interiors. This criterion states that shear-induced turbulence can be sustained provided 
\begin{gather}
    J\Pra < (J\Pra)_c,
    \label{eqn:modric}
\end{gather}
where $(J\Pra)_c$ is a constant he argues must be of order $10^{-3}$, and $\Pra=\nu/\kappa_T$ is the Prandtl number, defined as the ratio of the kinematic viscosity $\nu$ and the thermal diffusivity $\kappa_T$. 
Noting that $\Pra\sim10^{-5}-10^{-9}$ in typical stellar interiors \citep[][]{garaud_journey_2021}, Zahn's criterion suggests that flows with $J\sim10^2-10^6$ would be unstable, which is within the range of expected Richardson numbers in stars (although some have much larger Richardson numbers still). 
\cite{zahn_circulation_1992} later derived a turbulent mixing coefficient resulting from these diffusive stratified shear instabilities, which can in principle be used to model both chemical transport ($D_{\rm turb}$) or momentum transport $(\nu_{\rm turb})$: 
\begin{equation}
    D_{\rm turb} \simeq \nu_{\rm turb} = C\frac{\kappa_T}{J},
    \label{eq:ZahnDturb}
\end{equation}
where $C$ is a constant of order unity. 

The two components of Zahn's turbulent mixing prescription, namely equations (\ref{eqn:modric}) and (\ref{eq:ZahnDturb}), have been tested against direct numerical simulations, and found to be valid, in certain limits, for {\it non-rotating} diffusive shear flows \citep[see in particular][]{prat_mixing_2016,garaud_turbulent_2017}.
Both studies found that Zahn's stability criterion applies with $(J\Pra)_c\approx0.007$. They also found that (\ref{eq:ZahnDturb}) is correct for diffusive shear flows, but only as long as the turbulence is local, and $J\Pra \ll (J\Pra)_c$. A modified version of the mixing prescription that takes both issues into account was recently proposed by \cite{garaud_turbulent_2017}, and is discussed in more detail in Section \ref{sec:shearcomp}. Testing the validity of Zahn's model for rotating shear flows, which was its intended purpose (and the way it is commonly used in stellar evolution codes) is one of the goals of this paper.

Finally, note that a key aspect of diffusive shear instabilities is that they are not linearly excited when $J \gg 1$, but instead, emerge through nonlinear (subcritical) instabilities \citep[see, e.g.][]{garaud_stability_2015,garaud_turbulent_2016}. As such, they are subject to hysteresis \citep[see][]{gagnier_turbulent_2018}, and are only excited in this subcritical regime provided a minimum amount of turbulence is already present in the system to ``prime'' the instability. This, to our knowledge, is not accounted for in any stellar evolution code.

\subsection{Adding rotation: the GSF instability}

Although Zahn himself suggested that the results he derived for non-rotating shear flows might be directly applied to flows with rotational shear \citep{zahn_circulation_1992}, it is not obvious as there are several complications caused by rotation. First, it can influence the dynamics of the shear instability itself, by constraining the turbulence to be progressively more invariant along the rotation axis as the rotation rate increases \citep{gallet_2015}. Second, it is known to drive centrifugal instabilities, that ultimately extract energy from the angular momentum gradient rather than from the shear, and are therefore distinct from shear instabilities  \citep{rayleigh_dynamics_1917,solberg_mouvement_1936,hoiland_interpretation_1939,hoiland1941avhandliger}. In the absence of viscosity and thermal diffusion, 
and in the equatorial region of a star (which we are concerned with in this paper), the so-called Solberg-H\o iland criterion for centrifugal instability in differentially rotating, stratified shear flows, reads
\begin{gather}
   \frac{1}{r^3} \frac{\partial }{\partial r} (r^4\Omega^2) + N^2 = 2\Omega(2\Omega+S_\Omega) + N^2 \leq 0 
\label{eq:socrit}
\end{gather}
in the notation introduced above. 
In the absence of stratification ($N^2 = 0$), this criterion recovers the well-known Rayleigh instability criterion for centrifugal instabilities \citep{rayleigh_dynamics_1917}, which states that angular momentum must decrease outward for instability to occur. As such, the instability can only be excited when $S_\Omega$ is sufficiently negative. In a radiative zone ($N^2 > 0$), the thermal stratification acts to stabilize the flow as expected. In most stars, this would be sufficient to suppress centrifugal instabilities altogether, were it not -- as in the case of shear instabilities -- for non-adiabatic effects.  

Indeed,  \cite{goldreich_differential_1967} and \cite{fricke_instabilitat_1968} demonstrated that taking into account the effects of thermal diffusion greatly relaxes the instability criterion, which now reads
\begin{gather}
    \frac{1}{r^3} \frac{\partial }{\partial r} (r^4\Omega^2) + \frac{\nu}{\kappa_T}N^2 = 2\Omega(2\Omega+S_\Omega) + \Pra N^2 \leq 0, 
    \label{eqn:gsfdim}
\end{gather}
in the equatorial region of a star. Since $\Pra \ll 1$, this criterion is much more easily satisfied than (\ref{eq:socrit}). 
The instabilities that arise in that case are now known as GSF instabilities.  

\cite{goldreich_differential_1967} immediately noticed the strong similarity between the GSF instability and the double-diffusive fingering instability \citep[see also][for a detailed comparison of the two]{barker_angular_2019}. Indeed, the fingering instability exists in fluids that have a stable thermal stratification and an unstable compositional stratification. On sufficiently small scales, thermal diffusion reduces the stabilizing role of the temperature gradient, enabling perturbations to draw energy from the unstable composition gradient. Similarly in the case of the GSF instability, thermal diffusion enables small-scale perturbations to extract energy from the unstable angular momentum gradient. In both cases a sensible scale for the instability is
\begin{equation}
    d = \left( \frac{\kappa_T\nu}{N^2}\right)^{1/4}.
\end{equation}
where we note that the fastest-growing modes can take a larger or smaller value than $d$ depending on the parameters \citep[see e.g. the Appendix of][]{barker_angular_2020}. The unstable region of parameter space can be written as 
\begin{equation}
    1 \le R_0 \le \frac{\kappa_U}{\kappa_T},
    \label{eq:instabscrit}
\end{equation}
where $\kappa_U$ is the diffusivity associated with the unstable field (i.e. the compositional diffusivity in the fingering case, and the kinematic viscosity in the GSF case), and $R_0$ is a non-dimensional ratio of the square of timescales associated with the stabilizing and destabilizing stratifying components, respectively:
\begin{eqnarray}
    R_0 = -\frac{N^2}{N_C^2} \mbox{    in the fingering case,} \\
    \label{eq:R0critfingering}
    R_0 = -\frac{N^2}{2\Omega(2\Omega+S_\Omega)} \mbox{    in the GSF case,} \label{eq:R0critGSF}
\end{eqnarray}
where $N_C^2$ is the square of the buoyancy frequency associated with the composition gradient (which is negative when the latter is unstably stratified), and $2\Omega(2\Omega+S_\Omega)$ is the square of the epicyclic frequency associated with the angular momentum gradient (which is again negative when the latter is unstably stratified). In fact, the similarity between the two systems is so strong that they are exactly analogous in two dimensions, under an appropriate change of variables \citep[see a nice exposition of this analogy by][]{barker_angular_2019}. As a result, many of the results recently obtained in the fingering context  \citep[see the reviews by][]{garaud_double-diffusive_2018,Garaud_DCreview_2020} apply in the GSF case. Notably, it is possible to derive approximate analytical expressions for the growth rate of the GSF instability at low Prandtl number \citep[following the work of ][]{brown_chemical_2013}, and to construct a model for turbulent momentum transport that fits the data from direct numerical simulations remarkably well for a wide range of parameters \citep{barker_angular_2019}. This model can be written as 
\begin{equation}
   \nu_{\rm turb} = \frac{C_{B}^2}{S_\Omega} \frac{S_\Omega + 2 \Omega}{ \lambda  + \nu k^2 } \frac{\lambda^2}{k^2}, \label{eq:nuturbbarkerdim}
\end{equation}
when (\ref{eq:instabscrit}) is satisfied, where $\lambda$ and $k$ are the growth rate and wavenumber of the fastest-growing GSF modes, respectively, and $C_{B}$ is a universal constant that is fitted to the data (see more on this in Section \ref{sec:barkercomp}). The constant $C_B$ is related to the constant $A$ in \cite{barker_angular_2019}, see their equation (31), as $C_B^2 = A^2/2$.

\subsection{Coexistence of shear and GSF instabilities}

From the respective criteria for diffusive shear instabilities and GSF instabilities, we see that an interesting situation can arise when $S_\Omega < 0$, in which both instabilities are excited at the same time. In stellar evolution codes such as MESA \citep{Paxton_MESAcode_2011}, this situation is usually dealt with by computing a mixing coefficient for each instability, and adding them together to obtain a ``total'' mixing coefficient. However, this general practice is considered heretical by most fluid dynamicists, as there is a wealth of evidence showing that doing this often gives nonsensical results. For example, adding shear to convection or fingering convection can reduce mixing considerably  \citep{Garaud_shear_finger_2019,Blass_shear_convection_2021} instead of increasing it. There are also well-known cases in which two processes that would normally be stable, when taken individually, become unstable when they interact \citep{Hughes_Weiss_1995,Radko_thermoshear_2016}.

Based on these examples, one may naturally ask the question of what {\it really} happens when shear instabilities coexist with GSF instabilities. Of course, one may argue that by contrast with the examples cited above, the presence of shear is accounted for in the GSF instability criterion. However, the latter is a local model, that ignores the possibility of global shearing modes. In addition, it also ignores the contribution of the subcritical branch of diffusive shear instabilities. Conversely, the shear instability model ignores rotation entirely, but the latter could affect both the instability criterion, and the mixing model. As such it is important to revisit the problem of diffusive instabilities in rotating, stratified shear flows, and see whether the existing mixing models described above apply or not.





We begin by presenting the model setup used in this paper in Section \ref{sec:model}.
In Section \ref{sec:linstab}, we perform a linear stability analysis of the model system, and demonstrate the existence of coexisting instabilities in Section \ref{sec:linstabres}. In Section \ref{sec:numsim}, we present some qualitative results of the numerical investigation, while Section \ref{sec:numquant} studies them more quantitatively by comparing the measured mixing coefficients with the models of \cite{garaud_turbulent_2017} and \cite{barker_angular_2019}. This section will also reveal some of the more unusual aspects of the interaction between shear and GSF instabilities. Finally, Section \ref{sec:con} summarizes our results, and discusses their implications for stellar evolution models.

\section{Model Setup}
\label{sec:model}

\begin{figure}
    \centering
    \tdplotsetmaincoords{80}{230}
\begin{tikzpicture}[tdplot_main_coords,scale=0.5,line join=bevel]
\tikzmath{
\f = 3;
}
\coordinate (d) at (3,3,7);
\coordinate (C) at ({4*sin(66)*cos(30)},{4*sin(66)*sin(30)},-7);
\coordinate (O) at (0,0,0);

\tdplotsetcoord{P}{8}{66}{30}
    \draw[thick,dotted] (O) -- (Px);
    \draw[thick,dotted] (Px) -- (Pxz);
    \draw[thick,dotted] (Px) -- (Pxy);

    \draw [thick,dotted,domain={-99}:{80}, samples=100] plot ( {5*cos(\x)+4*sin(66)*cos(30)},{4*sin(66)*sin(30)},{5*sin(\x) - 7});
    \draw [thick,domain={81}:{260}, samples=100] plot ( {5*cos(\x)+4*sin(66)*cos(30)},{4*sin(66)*sin(30)},{5*sin(\x) - 7});
    \draw [thick,dashed,domain=0:{360}, samples=100] plot ( {4*cos(\x)+4*sin(66)*cos(30)},{4*sin(66)*sin(30)},{4*sin(\x) - 7});

    \draw [draw=red,domain=0:{4*cos(66)}, samples=100] plot ({4*sin(66)*cos(30)+sin(deg(2*pi*\x/8/cos(66)))},0,{4*cos(66)-\x}) -- ($ (Px) !0.5! (O) $) -- plot ({4*sin(66)*cos(30)+sin(deg(2*pi*\x/8/cos(66)))},{8*sin(66)*sin(30)},\x);
    \draw [draw=blue,domain={4*cos(66)}:{8*cos(66)}, samples=100] plot ({4*sin(66)*cos(30)+sin(deg(2*pi*\x/8/cos(66)))},{8*sin(66)*sin(30)},\x) -- ($ (Pxz) !0.5! (Pz) $) -- plot ({4*sin(66)*cos(30)+sin(deg(2*pi*\x/8/cos(66)))},0,{12*cos(66)-\x});

    \draw[thick] (O) -- (Py) -- (Pyz) -- (Pz) -- cycle;
    \draw[thick] (Pz) -- (Pyz) -- (P) -- (Pxz) -- cycle;
    \draw[thick] (Py) -- (Pxy) -- (P) -- (Pyz) -- cycle;

    \draw[thick,<-] ( $ (Pxz) !0.53! (O) $ ) -- ( $ (Pxz) !0.5! (O) + (-\f,-\f,-0.5) $ ) node (forcing) at ( $ (Pxz) !0.5! (O) + (-\f*1.1,-\f*1.1,-0.5) $ ) {$\mathbf F$};

    \fill ( $(0,0,4) + (C)$ ) circle (0.1cm);
    \draw[thin] (Pxy) -- ( $(0,0,4) + (C)$ ) -- (O);

    \draw[very thick] (C) circle (5cm);
    \draw[thick,->] ( $(C) + (0,5,0)$ ) -- ( $(C) + (0,7,0)$ ) node (omega) at ( $(C) + (0,7.7,0)$ ) {$\mathbf \Omega$};
\draw[thick,->] ( $(d)  + (C)$ ) -- ($(d) + (1.5,0,0) + (C) $) node[anchor=south east]{$x$};
\draw[thick,->] ( $(d)  + (C)$ ) -- ($(d) + (0,1.5,0) + (C) $) node[anchor=north east]{$y$};
\draw[thick,->] ( $(d)  + (C)$ ) -- ($(d) + (0,0,1.5) + (C) $) node[anchor=south]{$z$};
\end{tikzpicture}
    \caption{Model geometry. An imposed sinusoidal body force ${\bf F}$ drives an equatorial mean flow in the azimuthal direction, that varies with $z$ but is invariant in $y$.}
    \label{fig:model}
\end{figure}
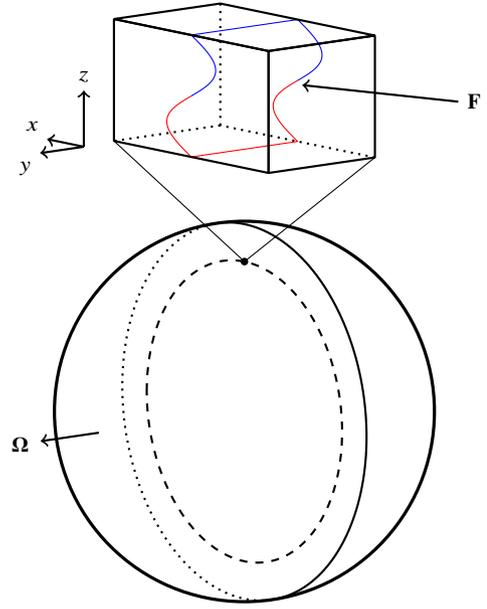

We consider a Cartesian domain located at the equator of a star, rotating with a constant angular velocity ${\bf \Omega}=\Omega_0 \unitvect_z$. The unit vectors $(\unitvect_x,\unitvect_y,\unitvect_z)$ are chosen such that $\unitvect_y$ points in the direction of ${\bf \Omega}$, $\unitvect_z$ points in the direction of $-{\bf g}$ (where ${\bf g}$ is gravity), and $\unitvect_x$ is in the azimuthal direction and is chosen so that the system right-handed (see Figure \ref{fig:model}).

 The domain is assumed to be located in a radiative zone, which is therefore stably stratified in terms of the potential temperature. We ignore for simplicity the possibility of a compositional stratification.
Assuming that the domain size is smaller than any of the local scaleheights, we use the Boussinesq approximation \citep{spiegel_boussinesq_1960}. Consistent with this approximation, we use a linear background thermal stratification in the $z$ direction as $T_0(z)=T_m + zT_{0z}$, where $T_m$ and $T_{0z}$ are constant.  Perturbations to this background are assumed to be periodic in all three directions. Finally, we assume the presence of a body force of the form ${\bf F}=F_0\sin(k_s z)\unitvect_x$, of amplitude $F_0$ and wavenumber $k_s$, that drives an azimuthal flow (see Figure \ref{fig:model}).

The following dimensional equations govern the dynamics of the model described above:
\begin{gather}
    \alignedb
    &\rho_m\left(\frac{\partial {\bf u}}{\partial t}+{\bf u} \cdot \nabla {\bf u}+ 2{\bf \Omega}\times {\bf u}\right) = -\nabla p + \rho {\bf g} + \rho_m \nu \nabla^2 {\bf u} + F_0\sin(k_s z)\unitvect_x, \\
    &\frac{\partial T}{\partial t}+{\bf u}\cdot\nabla T + w\left( T_{0z}-\frac{d T_{\rm ad}}{d z}\right) = \kappa_T\nabla^2 T, \\
    &\nabla\cdot{\bf u} = 0, \\
    &\frac{\rho}{\rho_m} = -\alpha T,
    \alignede
\end{gather}
where $\rho$, $p$ and $T$ are the density, pressure and temperature perturbations away from hydrostatic equilibrium, $\rho_m$ is the mean density of the region considered, and ${\bf u} = (u, v, w)$ is the velocity field. We assume the kinematic viscosity $\nu$, thermal diffusivity $\kappa_T$, the local gravity $g$, and the thermal expansion coefficient $\alpha=- \rho_m^{-1}(\partial \rho/\partial T)$ to be constant. The adiabatic temperature gradient is  $dT_{\rm ad}/dz = -g/c_p$ where $c_p$ is the specific heat at constant pressure.

Because of the imposed sinusoidal forcing, the system has a laminar steady state solution given by:
\begin{gather}
    {\bf u}_L = \frac{F_0}{\rho_m\nu  k_s^2}\sin(k_s z)\unitvect_x.
    \label{eqn:lamflow}
\end{gather}
We can non-dimensionalize the governing equations using the amplitude of the laminar solution $U_L=F_0/(\rho_m\nu k_s^2)$ as a velocity scale and the spatial scale of the laminar solution $k_s^{-1}$ as a lengthscale (and thus define $t_s=(k_sU_L)^{-1}$ as a time scale). The corresponding non-dimensional equations are:

\begin{gather}
    \alignedb
    &\frac{{\rm D}\hat{\bf u}}{{\rm D}\hat t} = -\nabla \hat p + \Ric \hat T\hat {\bf e}_z + \frac{1}{\Rey}\nabla^2 \hat{\bf u} - \frac{\hat w}{\Ros}\unitvect_x + \frac{\hat u}{\Ros}\unitvect_z +  \frac{1}{\Rey} \sin \hat z \unitvect_x, \\
    &\frac{\partial \hat T}{\partial \hat t} + \hat{\bf u}\cdot\nabla \hat T + \hat w = \frac{1}{\Pec}\nabla^2 \hat T, \\
    & \nabla\cdot \hat{\bf u} = 0,
    \alignede
    \label{eqn:nondim-lam}
\end{gather}
where hats denote non-dimensional variables (and  the gradient operators are now implicitly non-dimensional), with the following dimensionless parameters:
\begin{gather}
    \alignedb
    \Rey &= \frac{U_L}{ k_s\nu} = \frac{F_0}{\rho_m \nu^2 k_s^3}, \quad 
    \Ric =\frac{N^2}{U_L^2k_s^2}  = \frac{\alpha gT_{0z} \rho_m^2\nu^2k_s^2}{F_0^2}, \\
     \Pec &= \frac{U_L}{ k_s\kappa_T} = \frac{F_0}{\rho_m \kappa_T \nu  k_s^3} = \Pra \Rey, \quad 
    \Ros =\frac{ k_sU_L}{2\Omega_0} = \frac{F_0}{2\Omega_0 \rho_m\nu  k_s}.
    \alignede
    \label{eqn:nondimparams-lam}
\end{gather}
The Reynolds number $\Rey$ quantifies the ratio of the laminar flow shearing rate to the viscous diffusion rate across a lengthscale $k_s^{-1}$. The Richardson number $\Ric$ is the square of the ratio of the buoyancy frequency to the laminar flow shearing rate, which is used as a proxy for quantifying the potential energy lost in mixing the stratification to the kinetic energy gained from the shear. The P\'eclet number $\Pec$ is the ratio of the laminar flow shearing rate to the thermal diffusion rate. Finally, the Rossby number $\Ros$ is the ratio of the laminar flow shearing rate to the rotation rate, which measures the relative importance of the inertial terms and the Coriolis force.

\section{Linear stability analysis}
\label{sec:linstab}

\subsection{Linearized equations}

In this section, we look at the stability of the laminar steady state solution. We consider 3D infinitesimal perturbations (denoted by the primes) such that $\hat {\mathbf{u}} = \hat {\mathbf{u}}_{\it L} + \hat {\mathbf{u}}'$, where $\hat u_{\it L}(\hat z) = \sin \hat z$. 
When expanded in component form, the linearization of the system of equations (\ref{eqn:nondim-lam}) results in :
\begin{gather}
    \alignedb
    &\frac{\partial  {\hat u}'}{\partial \hat t} + \hat w'\frac{d \hat u_L}{d\hat z}+\hat u_L\frac{\partial \hat u'}{\partial \hat x}= -\frac{\partial \hat p'}{\partial \hat x} + \frac{1}{\Rey}\nabla^2 { \hat u}' - \frac{ \hat w'}{\Ros}, \\
    &\frac{\partial{\hat v}'}{\partial \hat t} + \hat u_L\frac{\partial \hat v'}{\partial \hat x}= -\frac{\partial \hat p'}{\partial \hat y} + \frac{1}{\Rey}\nabla^2 {\hat  v}', \\
    &\frac{\partial{ \hat w}'}{\partial \hat t} +\hat u_L\frac{\partial \hat w'}{\partial \hat x}= -\frac{\partial \hat p'}{\partial \hat z} + \Ric  \hat T' + \frac{1}{\Rey}\nabla^2 {\hat  w}' + \frac{ \hat u'}{\Ros},\\
    &\frac{\partial  \hat T'}{\partial \hat t} + \hat u_L\frac{\partial \hat T'}{\partial \hat x} + \hat  w' = \frac{1}{\Pec}\nabla^2 \hat  T', \\
    &\frac{\partial \hat u'}{\partial \hat x} + \frac{\partial \hat v'}{\partial \hat y} + \frac{\partial \hat w'}{\partial \hat z} = 0.
    \alignede
    \label{eqn:lineqns}
\end{gather}
 where all primed quantities are assumed to be small. 

This system of partial differential equations (PDEs) has non-constant coefficients, since the function $\hat u_L$ and its derivative are functions of $\hat z$. Nevertheless, we can use the periodicity of $\hat u_L(\hat z)$ to  transform it into a system of linear algebraic equations. We do so first by assuming that the perturbations $\hat q'(\hat x,\hat y,\hat z,\hat t)$ (where $\hat q'$ can be either ${\bf \hat u}'$, $\hat T'$ or $\hat p'$) can be written as normal modes of the form:
\begin{gather}
    \hat q'(\hat x,\hat y,\hat z,\hat t)= \tilde q(\hat z)\exp(i \hat k_x \hat x+i \hat k_y \hat y+\hat \lambda \hat t),
\end{gather}
where $\hat k_x$ and $\hat k_y$ are the non-dimensional wavenumbers in $\hat x$ and $\hat y$ respectively, and $\hat \lambda$ is a non-dimensional complex growth rate. We  seek solutions of the same periodicity as $\hat u_L(\hat z)$, satisfying:
\begin{gather}
    \tilde q(\hat z)=\sum\limits_{n=-\infty}^\infty q_n\exp(in\hat z).
\end{gather}
In practice, we limit the sum to a finite number of modes, ranging from $n=-N$ to $n=N$. We found $N=20$ to be sufficient and used this value throughout.

Substituting these ans\"atze into the linearized equations, we obtain a system of algebraic equations for the Fourier coefficients $q_n$, for $n=-N$ to $n=N$ (with the convention that the coefficients of modes with $n<-N$ and $n>N$ are set to zero):
\begin{gather}
    \alignedb
    & \hat \lambda u_n+\frac{w_{n-1}}{2}+\frac{w_{n+1}}{2}+\frac{ \hat k_x u_{n-1}}{2}-\frac{\hat k_x u_{n+1}}{2}  \\
    & \quad =-i\hat k_xp_n-\Rey^{-1}|\hat{\mathbf{k}}_n|^2u_n-\Ros^{-1}w_n, \\
    & \hat \lambda v_n+\frac{\hat k_x v_{n-1}}{2}-\frac{ \hat k_x v_{n+1}}{2}=-i\hat k_y p_n-\Rey^{-1}|\hat{\mathbf{k}}_n|^2v_n, \\
    & \hat \lambda w_n+\frac{\hat k_xw_{n-1}}{2}-\frac{\hat k_x w_{n+1}}{2}=-inp_n+\Ric T_n-\Rey^{-1}|\hat{\mathbf{k}}_n|^2w_n+\Ros^{-1}u_n, \\
    & \hat \lambda T_n+\frac{\hat k_x T_{n-1}}{2}-\frac{ \hat k_x T_{n+1}}{2}+w_n=-\Pec^{-1}|\hat{\mathbf{k}}_n|^2T_n, \\
    &  \hat k_x u_n+\hat k_y v_n+nw_n=0,
    \alignede
\end{gather}
where $\hat{\mathbf{k}}_n=(\hat k_x,\hat k_y,n)$. The dimension of the system can be reduced by eliminating the pressure $p_n$ analytically, resulting in:
\begin{gather}
    \alignedb
    & \hat \lambda u_n+\frac{w_{n-1}}{2}+\frac{w_{n+1}}{2}+\frac{\hat k_x u_{n-1}}{2}-\frac{\hat k_xu_{n+1}}{2}=-\Rey^{-1}|\hat{\mathbf{k}}_n|^2u_n-\Ros^{-1}w_n, \\
    &\quad +\frac{\hat k_x}{\hat k_y}\left[ \hat \lambda v_n + \frac{\hat k_x v_{n-1}}{2} - \frac{\hat k_x v_{n+1}}{2} + \Rey^{-1}|\hat {\mathbf{k}}_n|^2v_n \right] \\
    & \hat \lambda w_n+\frac{\hat k_xw_{n-1}}{2}-\frac{\hat k_xw_{n+1}}{2}=\Ric T_n-\Rey^{-1}|\hat{\mathbf{k}}_n|^2w_n+\Ros^{-1}u_n\\
    &\quad +\frac{\hat k_z}{\hat k_y}\left[ \hat \lambda v_n + \frac{\hat k_x v_{n-1}}{2} - \frac{\hat k_x v_{n+1}}{2} + \Rey^{-1}|\hat {\mathbf{k}}_n|^2v_n \right] \\
    & \hat \lambda T_n+\frac{\hat k_xT_{n-1}}{2}-\frac{\hat k_xT_{n+1}}{2}+w_n=-\Pec^{-1}|\hat {\mathbf{k}}_n|^2T_n \\
    &  \hat k_x u_n+\hat k_y v_n+nw_n=0.
    \alignede
    \label{eqn:reducedalg}
\end{gather}
This now takes the form of a generalized eigenvalue problem, namely
\begin{gather}
    {\bf A}(\Rey,\Ric,\Pec,\Ros,\hat k_x,\hat k_y){\bf X}=\hat \lambda{\bf B X},
\end{gather}
where ${\bf X}=(u_{-N},\dots,u_N, v_{-N},\dots,v_N, w_{-N},\dots,w_N,T_{-N},\dots,T_N)$ is the solution vector for some finite $N$, and ${\bf A}$ and ${\bf B}$ are two $4(2N+1)\times4(2N+1)$ matrices. This problem can be solved numerically for $\hat \lambda$ for any given set of parameters  $(\Rey,\Pec,\Ric,\Ros)$, and selected horizontal wavenumbers $\hat k_x$ and $\hat k_y$.  Note that there are, by construction, $4(2N+1)$ possible eigenvalues and eigenvectors of this system. However, we only keep the solution for which the real part of $\hat \lambda$ is largest, and call it $\hat \lambda(\Rey,\Pec,\Ric,\Ros,\hat k_x,\hat k_y)$.

\subsection{Two remarkable limits}
\label{sub:twolims}
Although we generally need to solve the system numerically, two remarkable limits can be derived analytically from the system of equations (\ref{eqn:lineqns}). The first limit is obtained assuming invariance of the perturbations in the
$y$-direction, so $\partial/\partial \hat y = 0$. The incompressibility condition then reduces to:
\begin{gather}
    \frac{\partial \hat u'}{\partial \hat x} +  \frac{\partial \hat w'}{\partial \hat z} = 0.
    \label{eqn:incomp}
\end{gather}
By taking the $\hat z$-derivative of the $x$-momentum equation and the $\hat x$-derivative of the $z$-momentum equation, one arrives at the following two equations:
\begin{gather}
    \alignedb
    &\frac{\partial}{\partial \hat z}\left(\frac{\partial{ \hat u}'}{\partial \hat t} + \hat w'\frac{d\hat u_L}{d\hat z}+\hat u_L\frac{\partial \hat u'}{\partial \hat x}\right)= \frac{\partial}{\partial \hat z}\left(-\frac{\partial { \hat p'}}{\partial \hat x} + \frac{1}{\Rey}\nabla^2 { \hat u}'\right) - \frac{1}{\Ros}\frac{\partial \hat w'}{\partial \hat z}, \\
    &\frac{\partial}{\partial \hat x}\left(\frac{\partial{ \hat w}'}{\partial \hat t} + \hat u_L\frac{\partial \hat w'}{\partial \hat x}\right)= \frac{\partial}{\partial \hat x}\left(-\frac{\partial { \hat p'}}{\partial \hat z} + \Ric  \hat T' + \frac{1}{\Rey}\nabla^2 { \hat w}'\right) + \frac{1}{\Ros}\frac{\partial \hat u'}{\partial \hat x}, 
    \alignede
\end{gather}
which are the only remaining equations directly affected by rotation. If we subtract one from the other, and use the  incompressibility condition (\ref{eqn:incomp}), terms containing $\Ros$ disappear from the system to yield,
\begin{gather}
    \alignedb
    \frac{\partial}{\partial \hat t}\left(\frac{\partial \hat u'}{\partial \hat z}-\frac{\partial \hat w'}{\partial \hat x}\right) + \left(\frac{\partial}{\partial \hat z}\left(\hat u_L\frac{\partial \hat u'}{\partial \hat x}\right)-\hat u_L\frac{\partial^2 \hat w'}{\partial \hat x^2}\right)+\frac{\partial}{\partial \hat z}\left(\hat w'\frac{d \hat u_L}{d \hat z}\right) \\ 
    = -\Ric\frac{\partial \hat T'}{\partial \hat x} + \frac{1}{\Rey}\nabla^2\left(\frac{\partial \hat u'}{\partial \hat z}-\frac{\partial \hat w'}{\partial \hat x}\right).
    \alignede
\end{gather}

This shows that rotation has no effect on the linear evolution of $\hat y$-invariant perturbations in the equatorial regions of a star. Hence, we expect to recover the stability properties of standard diffusive stratified shear instabilities for $\hat k_y=0$ modes, regardless of the rotation rate \citep[see, e.g.][]{garaud_stability_2015}.

On the other hand, if we assume $x$-invariance ($\partial/\partial \hat x=0$), (\ref{eqn:lineqns}) reduces to the following set of PDEs:
\begin{gather}
    \alignedb
    &\frac{\partial{ \hat u}'}{\partial \hat t} + \hat w'\frac{d\hat u_L}{d\hat z}= \frac{1}{\Rey}\nabla^2 { \hat u}' - \frac{ \hat w'}{\Ros}, \\
    &\frac{\partial{ \hat v}'}{\partial \hat t} = -\frac{\partial { \hat p'}}{\partial \hat y} + \frac{1}{\Rey}\nabla^2 { \hat v}', \\
    &\frac{\partial{ \hat w}'}{\partial \hat t} = -\frac{\partial {\hat  p'}}{\partial \hat z} + \Ric \hat T' + \frac{1}{\Rey}\nabla^2 { \hat w}' + \frac{ \hat u'}{\Ros}, \\
    &\frac{\partial \hat T'}{\partial \hat t} +\hat w' = \frac{1}{\Pec}\nabla^2 \hat T', \\
    &\frac{\partial \hat v'}{\partial \hat y} + \frac{\partial \hat w'}{\partial \hat z} = 0.
    \alignede
\end{gather}
If, for the moment, we further assume that the vertical scale of the instability is much smaller than the characteristic lengthscale of the shear (which can be verified a posteriori), then we can approximate the shear to be locally constant\footnote{A more formal approach would involve using a JWKB approximation on the governing equations.} with a value $\hat S$. The system now has constant coefficients, and we can successively eliminate variables to arrive at the following equation:
\begin{gather}
   \hat D_{\bf u}^2\hat D_T\nabla^2 \hat w' = -\left[\frac{1}{\Ros}\hat D_T\left(\frac{1}{\Ros}+\hat S\right)+\Ric \hat D_{\bf  u}\right]\frac{\partial^2 \hat w'}{\partial \hat y^2},
    \label{eqn:xinvariant}
\end{gather}
where $\hat D$'s are shorthand notations for the differential operators:
\begin{gather}
    \hat D_{\bf u}= \frac{1}{\Rey}\nabla^2-\frac{\partial}{\partial \hat t},\quad \hat D_{T}= \frac{1}{\Pec}\nabla^2-\frac{\partial}{\partial \hat t}.
\end{gather}

We assume normal mode solutions of the form $\hat w'\propto\exp(i\hat k_y \hat y+i\hat k_z \hat z+\hat \lambda \hat t)$ and obtain the algebraic equation:
\begin{gather}
    \left(\frac{\hat K^2}{\Rey}+\hat \lambda\right)^2\left(\frac{\hat K^2}{\Pec}+\hat \lambda\right)=-\left(\frac{\hat K^2}{\Pec}+\hat \lambda\right)\left[\frac{1}{\Ros}\left(\frac{1}{\Ros}+\hat S\right)+\Ric\left(\frac{\hat K^2}{\Rey}+\lambda\right)\right]\frac{\hat k_y^2}{\hat K^2},
\end{gather}
where $\hat K^2=\hat k_y^2+\hat k_z^2$. This can be expanded to give a third order polynomial equation in $\hat \lambda$:
\begin{gather}
    \alignedb
    \hat \lambda^3& + a_2\hat \lambda^2 + a_1\hat \lambda + a_0=0, \\
    a_2&=\left(\frac{1}{\Pec}+\frac{2}{\Rey}\right)\hat K^2, \\
    a_1&=\left(\frac{2}{\Rey\Pec}+\frac{1}{\Rey^2}\right)\hat K^4+\Ric\frac{\hat k_y^2}{\hat K^2}+\frac{1}{\Ros}\left(\frac{1}{\Ros}+S\right)\frac{\hat k_y^2}{\hat K^2}, \\
    a_0&=\frac{\hat K^6}{\Rey^2\Pec}+\Ric\frac{\hat k_y^2}{\Rey}+\frac{\hat k_y^2}{\Ros\Rey}\left(\frac{1}{\Ros}+\hat S \right).
    \alignede
\end{gather}

The absence of any solution with positive real part (which is necessary for stability) can be established using the Routh-Hurwitz theorem. For a third order polynomial, the Routh-Hurwitz criterion for stability is satisfied if and only if $a_2,a_0>0$ and $a_2 a_1>a_0$ \citep{anagnost_elementary_1991}. Specifically, the condition $a_0>0$ gives the following inequality:
\begin{gather}
    \frac{1}{\Rey^2}\frac{\hat K^6}{\hat k_y^2}>-\frac{1}{\Ros}\left(\frac{1}{\Ros}+\hat S\right)-\Ric\Pra.
\end{gather}
For this to be true for any value of $\hat k_y,\hat k_z\neq0$, the RHS must be non-positive. In other words, the system is  stable to all possible modes provided $
    0\ge-\Ros^{-1}\left(\Ros^{-1}+\hat S\right)-\Ric\Pra $
and conversely, can be unstable to some modes provided 
\begin{gather}
    0 > \frac{1}{\Ros}\left(\frac{1}{\Ros}+\hat S\right)+ \Ric\Pra.
    \label{eqn:gsfnondim}
\end{gather}
This is the non-dimensional equivalent of the GSF instability criterion (\ref{eqn:gsfdim}). 

This detour has enabled us to identify analytically two distinct modes of instability: the first one is a pure shearing mode, with $\hat k_y  = 0$, that does not know about rotation; the second one has $\hat k_x = 0$, and is a standard two-dimensional GSF mode (see above). As we shall see below, one or the other of these two modes tends to dominate the linear stability of the system in almost all of the parameter space.

\section{Linear stability results}
\label{sec:linstabres}

\begin{figure*}
     \caption{Linear stability analysis results. From left to right: growth rate map (log color bar), $x$-wavenumber map (linear color bar) and $y$-wavenumber map (log color bar). White areas are regions that are linearly stable. The red line is the marginal stability boundary for the GSF instability. $\Rey=10000$ for all maps. From top to bottom: $\Pec=1000,10,0.1$. }
     \label{fig:linstab}
     \centering
     \includegraphics[width=\textwidth]{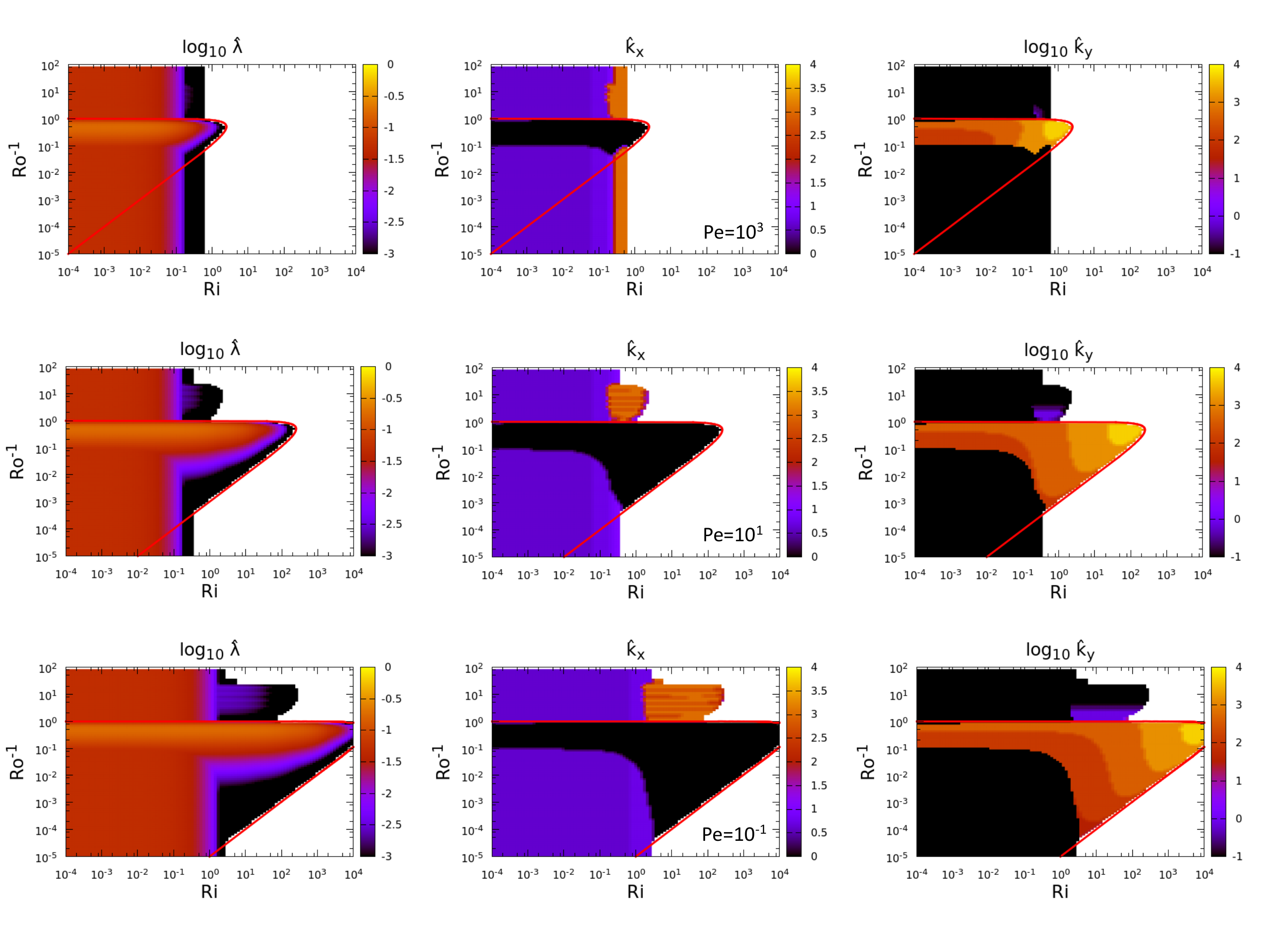}
\end{figure*}

 In this section, we now present and discuss the results of the linear stability analysis outlined in Section \ref{sec:linstab}. In what follows, we compute the fastest-growing perturbations to the stratified, rotating shear flow ${\bf \hat u}_L(\hat z)$, for a given set of model parameters $(\Rey,\Pec,\Ric,\Ros)$, by maximizing $\Re[\hat \lambda(\Rey,\Pec,\Ric,\Ros,\hat k_x,\hat k_y)]$ over all possible values of the horizontal wavenumbers $\hat k_x$ and $\hat k_y$. The maximization problem then returns the growth rate and wavenumbers of the fastest-growing modes. The latter are presented, for fixed $\Rey = 10000$ and varying $\Pec$, $\Ric$, and $\Ros$, in Figure \ref{fig:linstab}. 

Each row of Figure \ref{fig:linstab} corresponds to a particular value of $\Pec$, ranging from $1000$ (top row, weak thermal diffusion) down to $0.1$ (bottom row, strong thermal diffusion). In each row, the left-side panel shows the real part of the growth rate (denoted for simplicity with $\hat \lambda$), the middle panel shows the wavenumber $\hat k_x$, and the right-side panel shows the wavenumber $\hat k_y$, of the fastest-growing modes. Finally, within each panel, the quantity in question is shown as a color map, as a function of the Richardson number $\Ric$ (horizontal axis) and the inverse Rossby number $\Ros^{-1}$ (vertical axis). Stratification increases as $\Ric$ increases, and the rotation rate increases as $\Ros^{-1}$ increases (see equation \ref{eqn:nondimparams-lam}). If the system is linearly stable, the quantity is rendered using the white color.  

Overall, we clearly see the emergence of (at least) two distinct modes of instability, whose relative importance (in terms of which has the largest growth rate) depends on the input parameters. The first mode has $\hat k_y = 0$, $\hat k_x=  O(1)$, and a growth rate $\hat \lambda$ that appears to be independent of the rotation rate. The mode is stabilized if $\Ric$ exceeds a certain threshold. The second mode has $\hat k_x = 0$, a $\hat k_y$ that can be significantly greater than one, and dominates in a region of the $(\Ric,\Ros^{-1})$ plane whose shape is somewhat reminiscent of a plough. Based on the analysis of Section \ref{sub:twolims}, we see that the first mode is clearly a  shearing mode, while the second is clearly a GSF mode. This is also confirmed by plotting the marginal stability curve associated with the GSF instability in red (see equation \ref{eqn:gsfnondim}); the region below and to the left of the red curve is linearly unstable to GSF perturbations. We therefore conclude that there is a substantial range of parameters for which the GSF instability and the shear instability can in principle coexist, even if one dominates over the other from a linear perspective.  
Finally we also note that there appears to be a third, fully three-dimensional mode present for large rotation rate and intermediate stratification, that we will not discuss in this paper (as its physical interpretation is not well understood).

In the weakly rotating limit, ($\Ros^{-1} \rightarrow 0)$, the GSF mode is either stabilized or subdominant, and the shearing mode dominates. The growth rate of the shear instability tends to a constant of order 0.1 in the limit of weak stratification, and drops to zero as $\Ric$ exceeds a certain threshold whose value depends on the P\'eclet number. We can see from Figure \ref{fig:linstab} that the neutral stability line lies close to one at large P\'eclet number. This may seem surprising at first given that the Richardson criterion states that stratified shear flows are linearly stable if the gradient Richardson number, which is equal to $J=\Ric / \cos^2(\hat z)$ in this problem, is greater than 1/4 everywhere in the flow, which happens as soon as $\Ric > 1/4$. This apparent discrepancy can be resolved by noting that the standard Richardson criterion neglects viscous effects, and thus fails to capture viscous instabilities. Between $\Ric = 1/4$ and $\Ric = 1$, viscous modes exist and can be distinguished by their small growth rates. This has been shown by \cite{balmforth_stratified_2002} in the case of stratified 2D sinusoidal shear flows and by \cite{garaud_stability_2015} in the case of stratified 3D sinusoidal shear flows.

As $\Pec$ decreases below unity, the critical value of the Richardson number for shear instability $\Ric_c$ increases, consistent with the results of \cite{garaud_stability_2015}, who demonstrated that the marginal stability criterion satisfies $\Ric_c = O(\Pec^{-1})$ for a low P\'eclet number sinusoidal flow. Since the inverse of the P\'eclet number represents how thermally diffusive the system is, decreasing $\Pec$ is equivalent to increasing thermal diffusion in the system. Faster thermal diffusion acts as a destabilizing agent against the density stratification, allowing for the existence of unstable modes at higher values of $\Ric$.

Finally, it is worth remembering that the shear instability has a subcritical (nonlinear) branch (see Section \ref{sec:intro}). This subcritical instability does not appear in this linear stability figure, but can be excited by finite-amplitude perturbations of the right shape provided the product of the local Richardson number and the Prandtl number is lower than approximately $0.007$ \citep[at least in the non-rotating case, see][]{zahn_rotational_1974,prat_mixing_2016,garaud_turbulent_2017}. Since $\Pra = 0.1$ in the top row, $0.001$ in the middle row, and $0.00001$ in the bottom row, this nonlinear branch is irrelevant in the top row, but would exist up to $\Ric \simeq 7$ and $\Ric \simeq 700$ in the middle and bottom rows, respectively. 



At higher rotation rates (larger $\Ros^{-1}$), the GSF mode emerges, and dominates in the plough-shaped region described earlier. For large $\Pec$ (top row), this region has a fairly limited extent, but expands as $\Pec$ decreases (middle and bottom rows) to encompass higher values of the stratification and lower rotation rates, consistent with the criterion in equation (\ref{eqn:gsfnondim}), red line. 
Within the GSF region, we see that the $y$-wavenumber of the fastest growing GSF mode tends to increase as $\Ric$ increases. This can be understood noting that GSF instabilities are analogous to double diffusive fingering instabilities, whose characteristic wavenumber is proportional to $\Ray^{1/4}$ in the non-dimensionalization used here, where $\Ray=\Pec\Rey\Ric$ is the Rayleigh number. At constant $\Rey$ and $\Pec$, an increase in $\Ric$ therefore implies an increase in the characteristic wavenumber of the fastest growing mode.

In summary, we see that this model system has the potential for a wide range of interesting dynamics, raising many questions that are of relevance to mixing in stars, and of theoretical interest in fluid dynamics. In particular, it will be interesting to determine (1) Which linear mode of instability ends up dominating the system dynamics in the regions of parameter space where both coexist, and (2) What happens when the GSF coexists with the subcritical branch of the shear instability. 



\section{Numerical simulations}
\label{sec:numsim}

In what follows, we now turn to direct numerical simulations (DNS) to study the nonlinear aspects of instabilities of rotating diffusive stratified shear flows. We begin by briefly describing the algorithm and model setup used in the numerical experiments, before presenting the evolution of a characteristic simulation.

\subsection{Numerical code: PADDI}

 The numerical experiments presented in this work were performed using the PADDI code. PADDI is a high-performance pseudo-spectral code originally developed to solve double-diffusive hydrodynamic equations over a triply-periodic 3D domain \citep{stellmach_dynamics_2011}.  Salient aspects of the code are presented in \cite{traxler_numerically_2011}. The original code was modified to include both the Coriolis force \citep{moll_effect_2017} and a sinusoidal body force \citep{garaud_stability_2015} to suit the needs of the study.

\subsection{Forcing-based non-dimensionalization}

For the purpose of the linear stability analysis presented in Sections \ref{sec:linstab} and \ref{sec:linstabres}, we employed a non-dimensionalization based on the amplitude of the laminar steady state solution, $U_L=F_0/(\rho_m\nu  k_s^2)$. This non-dimensionalization is useful when looking at the early stages of development of the instability starting from the laminar solution, but is not appropriate once the flow becomes nonlinear, and the mean shear decreases as a result of momentum transport by the turbulence. \citet{garaud_turbulent_2016} proposed that a more relevant non-dimensionalization in turbulent, non-rotating, shear flows can be derived from assuming a balance between inertial terms and the forcing in the momentum equation,   
such that:
\begin{equation}
    \rho_m({\bf u}\cdot\nabla{\bf u})\cdot\unitvect_x\sim F_0.
\end{equation}
Then we can define a new flow amplitude $U_F$ as:
\begin{equation}
    U_F=\left(\frac{F_0}{k_s\rho_m}\right)^{1/2}.
\end{equation}
This velocity scale is independent of any diffusivity, and thus ought to be more relevant than $U_L$ once turbulence has fully developed. 

Interestingly, the quantities $U_F$ and $U_L$, respectively, can now be seen as estimates for the minimum and maximum possible amplitudes of the body-forced mean flow achievable in a statistically stationary state for the selected model setup. Indeed, the amplitude of the mean flow is expected to be largest in the absence of any turbulent dissipation, where it takes the value $U_L$, while it is expected to be smallest when the turbulent viscosity is largest. Hidden in the dimensional argument above is the assumption that the turbulent eddies have (1) a lengthscale of order $k_s^{-1}$ and (2) a characteristic velocity of the order of the mean flow velocity. These are the largest possible length- and velocity scales for this flow, so the corresponding turbulent viscosity implied in this argument is also the largest one achievable. Hence $U_F$ is an estimate of the smallest mean flow amplitude achievable in this setup.

Using a new system of units where the velocities are scaled by $U_F$ instead of $U_L$ (and the unit timescale is correspondingly changed to $1/k_s U_F$), we obtain a set of non-dimensional equations that looks almost identical to the system (\ref{eqn:nondim-lam}):
\begin{gather}
    \alignedb
    &\frac{{\rm D}\check{\bf u}}{{\rm D}\check t} = -\nabla \check p + \Ric_F \check T\hat {\bf e}_z + \frac{1}{\Rey_F}\nabla^2 \check{\bf u} - \frac{\check w}{\Ros_F}\unitvect_x + \frac{\check u}{\Ros_F}\unitvect_z +  \sin \check z \unitvect_x, \\
    &\frac{\partial \check T}{\partial \check t} + \check{\bf u}\cdot\nabla \check T + \check w = \frac{1}{\Pec_F}\nabla^2 \check T, \\
    & \nabla\cdot \check{\bf u} = 0,
    \alignede
    \label{eqn:nondim-for}
\end{gather}
where $\check {\bf u} = {\bf u}/U_F$, $\check t = t U_F k_s$,  $\check T = \hat T$, and where:
\begin{gather}
    \alignedb
    \Rey_F &= \frac{U_F}{k_s\nu} = \left(\frac{F_0}{\rho_m\nu^2 k_s^3}\right)^{1/2}=\Rey^{1/2}, \\
    \Ric_F &=\frac{N^2}{U_F^2 k_s^2} = \frac{N^2\rho_m}{k_sF_0}=\Rey\Ric, \\
    \Pec_F &=\frac{U_F}{ k_s\kappa_T} = \left(\frac{F_0}{\rho_m\nu  k_s^3\kappa_T^2}\right)^{1/2}=\Rey^{-1/2}\Pec, \\
    \Ros_F &= \frac{k_s U_F}{2\Omega_0} = \left(\frac{ k_sF_0}{4\Omega_0^2 \rho_m}\right)^{1/2}=\Rey^{-1/2}\Ros.
    \alignede
    \label{eqn:nondimparams-for}
\end{gather}
The numerical results described in the following sections will be given in this new non-dimensionalization. Note that since the unit lengthscale has not changed, $\hat x = \check x$, and similarly for $\check y$ and $\check z$.
\cite{garaud_turbulent_2016} demonstrated that $\Rey_F$,$\Pec_F$ and $\Ric_F$ are relatively good estimates for the actual turbulent Reynolds, P\'eclet and Richardson numbers in {\it non-rotating} stratified shear flows driven by a body force. This is not necessarily true anymore in the rotating case, but this system of units is still more appropriate than the one based on the laminar flow.  

\subsection{Characteristic simulation output}
\label{sec:charsim}

We begin by presenting the evolution of a characteristic simulation, whose governing parameters are $\Rey_F=100,\Pec_F=0.1,\Ric_F=1000$, and $\Ros_F^{-1}=1$, or equivalently in the non-dimensionalization based on the laminar flow, $\Rey=10000,\Pec=10,\Ric=0.1$, and $\Ros^{-1}=0.01$. According to the linear stability analysis of the previous section, the fastest-growing mode of instability starting from the laminar equilibrium state should be a shearing mode. The non-dimensional numerical domain size is $(\check L_x = 4\pi,\check L_y= 2\pi,\check L_z= 2\pi)$, which is sufficiently long to allow for the natural development of the basic shear instability without constraining the flow too much. Since the GSF modes are smaller-scale, this domain size ought to be sufficient in the GSF limit as well. The numerical resolution selected for this simulation has $384\times192\times192$ equivalent grid points.
 
\begin{figure*}
     \caption{Characteristic simulation with governing parameters $\Rey_F=100,\Pec_F=0.1,\Ric_F=1000$, and $\Ros_F^{-1}=1$, or equivalently in the non-dimensionalization based on the laminar flow, $\Rey=10000,\Pec=10,\Ric=0.1$, and $\Ros^{-1}=0.01$. Panel (a): time evolution of $\check u_{\rm rms}$, $\check v_{\rm rms}$ and $\check w_{\rm rms}$. 
     Panel (b): snapshots of the $z$-profiles of the mean flow $\check {\bar u}$. The times correspond to the dotted gray lines in (a). 
     Panel (c): a representative snapshot of $\check w$ taken at $\check t = 42.5$. The axes are such that $x$ is along the long direction of the domain, and $z$ points upwards. 
     Panel (d): the trajectory of the simulation (thin gray line) and steady state (thick cyan line) plotted over the linear stability map in $\log_{10}\hat k_y$. It lies between two triangles which denote the laminar limit (upwards triangle) and the fully turbulent limit (inverted triangle). The simulation is initialized in the linearly stable region and ends up in a statistically steady state in the GSF-dominated region. The dashed solid line shows the extent of the subcritical branch of diffusive shear instability (assuming it is not affected by rotation).}
     \centering
     \includegraphics[width=0.8\textwidth]{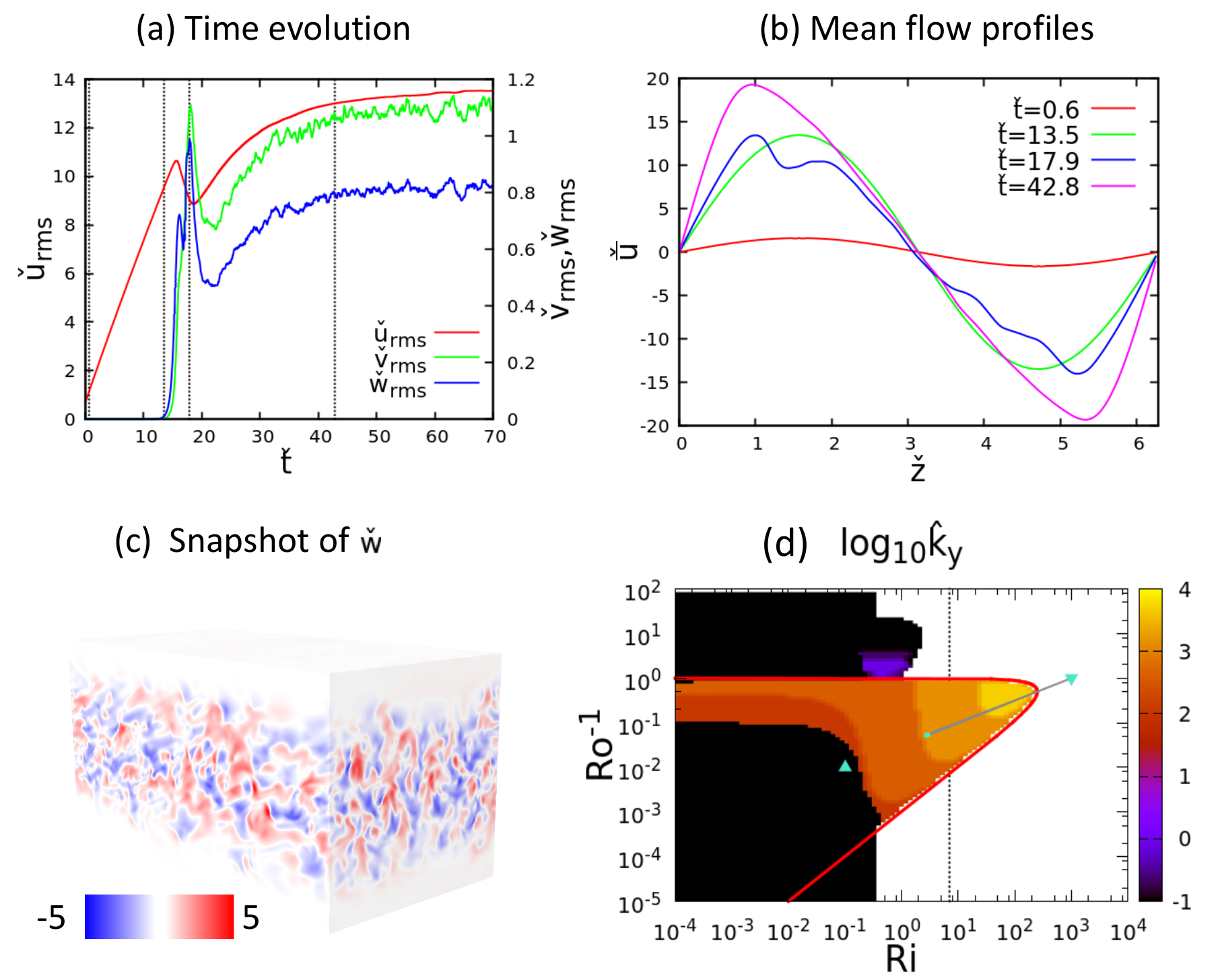}
     \label{fig:plotshowcase}
\end{figure*}

 The simulation is initialized with a sinusoidal streamwise flow profile given by $\check u_0(\check x, \check y, \check z, 0) = \sin (\check z)$, plus small random perturbations that seed the instabilities. We note that this flow is {\it not} the laminar equilibrium solution for the system (which would have an amplitude of $\Rey_F$ in this non-dimensionalization). Instead, the flow is stable at time $\check t = 0$. The evolution of various quantities of interest with time is presented in Figure \ref{fig:plotshowcase}.

Figure \ref{fig:plotshowcase}(a) shows the evolution of $\check u_{\rm rms}(\check t)$, $\check v_{\rm rms}(\check t)$ and $\check w_{\rm rms}(\check t)$, where 
\begin{equation}
\check u_{\rm rms}(\check t) = \left( \frac{1}{\check L_x \check L_y \check L_z} \int_D \check u^2( \check {\bf x},\check t) d^3\check {\bf x}\right)^{1/2}, 
\end{equation}
(and similarly for $\check v$ and $\check w$). The constant forcing in the streamwise direction causes the amplitude of the sinusoidal flow (and therefore $\check u_{\rm rms}(\check t)$) to increase linearly with time until about $\check t = 15$, where it reaches the linear instability threshold  for GSF modes to grow. At this point, perturbations begin to grow exponentially (see, e.g. $\check v_{\rm rms}(\check t)$ and $\check w_{\rm rms}(\check t)$). Visual inspection of $\check w$ shows that the perturbations are limited spatially to the region where the mean shear is negative, and are initially invariant in $\check x$, which is as expected from the GSF instability. Eventually, the perturbations begin to affect the mean flow and nonlinear saturation occurs around $\check t = 20$. At this point the turbulence is still limited to regions of negative shear, but has become fully three dimensional, as illustrated in the snapshot of $\check w$ taken at $\check t = 43$ (see Figure \ref{fig:plotshowcase}(c)). The shear in the laminar regions continues to grow slowly on a viscous timescale in response to the imposed forcing, until the system finally reaches a statistically stationary state  around $\check t=70$. 

Figure \ref{fig:plotshowcase}(b) shows the mean streamwise flow profile \begin{equation}
    {\check {\bar u}}(\check z,\check t) = \frac{1}{\check L_x \check L_y} \int_0^{\check L_x}\int_0^{\check L_y} \check u(\check x,\check y,\check z,\check t) d\check x d\check y 
\end{equation}
at selected times. It has a perfect sinusoidal shape at early times, but acquires a marked asymmetry once the GSF instability develops, whereby the turbulent region for $1 < \check z < 5$ has a weaker shear, while the laminar regions for $0 < \check z < 1$ and $ 5 < \check z < 2\pi$ have a much larger shear. To understand why the mean flow becomes strongly asymmetric, note that the horizontal average of the momentum equation is 
\begin{gather}
    \frac{\partial {\check {\bar u}}}{\partial \check t} + \frac{\partial }{\partial \check z}(\overline{\check u\check w}) = \frac{1}{\Rey_F} \frac{\partial^2 {\check {\bar u}}}{\partial \check z^2} + \sin(\check z),
    \label{eq:momaverage}
\end{gather}
where $\overline{\check u\check w}$ is the Reynolds stress. Assuming that this turbulent stress behaves in a diffusive manner, we can define a turbulent diffusivity $\check \nu_{\rm turb}$ in the usual way, 
\begin{gather}
    \overline{\check u\check w} = -\check \nu_{\rm turb}\frac{\partial {\check {\bar u}}}{\partial \check z},
    \label{eqn:nu}
\end{gather}
in which case (\ref{eq:momaverage}) becomes:

\begin{gather}
    \frac{\partial {\check {\bar u}}}{\partial \check t} = \frac{\partial}{\partial \check z}\left[(\Rey_F^{-1}+\check \nu_{\rm turb})\frac{\partial {\check {\bar u}}}{\partial \check z}\right] + \sin(\check z).
\end{gather}

In a statistically stationary state, the time derivative can be ignored, and integration in $\check z$ yields:
\begin{gather}
     \frac{\partial {\check {\bar u}}}{\partial \check z} = \frac{\cos(\check z)}{ \Rey_F^{-1}+\check \nu_{\rm turb}} .
\end{gather}
This shows that the amplitude of the mean shear $\partial {\check {\bar u}}/\partial \check z$ must be weaker in turbulent regions  (assuming $\check \nu_{\rm turb}>0$), and stronger in laminar regions where $\hat \nu_{\rm turb}\simeq 0$).



To understand why the system ends up being governed by the GSF instability rather than the shear instability, we compute the ``trajectory'' of the simulation on the $(\Ric,\Ros^{-1})$ stability diagram. To do so, note that if we allow ourselves to approximate the mean flow profile by the formula ${\check {\bar u}}(\check z,\check t) \simeq \check A(\check t) \sin(\check z)$, with $\check A(\check t)$ varying slowly with time, then we can use the frozen-in approximation to perform a linear stability of this flow at any time $\check t$. This is easily done by solving the linear system  (\ref{eqn:reducedalg}), with $(\Rey,\Pec,\Ric,\Ros)$ replaced by effective parameters  $(\Rey_{\rm eff},\Pec_{\rm eff},\Ric_{\rm eff},\Ros_{\rm eff})$, where 
\begin{eqnarray}
\Rey_{\rm eff} = \frac{\check A U_F}{\nu k_s} = \check A \Rey_F \\
\Pec_{\rm eff} = \frac{\check A U_F}{\kappa_T k_s} = \check A \Pec_F \\
\Ric_{\rm eff} = \frac{N^2}{\check A^2 U_F^2 k_s^2} = \frac{\Ric_F}{\check A^2} \\
\Ros_{\rm eff} = \frac{k_s \check A U_F}{2 \Omega_0} = \check A \Ros_F. 
\end{eqnarray}
We can verify that this is indeed consistent: the laminar steady-state solution has amplitude $\check A = \Rey_F$, and in that case $\Rey_{\rm eff} = \Rey_F^2 = \Rey$ (and similarly for the other parameters), so we indeed recover (\ref{eqn:reducedalg}). With $\check A = 1$, by contrast, $\Rey_{\rm eff} = \Rey_F$ (and similarly for the other parameters), which gives an estimate of the stability of the system in its ``weakest shear'' configuration. In practice, we obtain a quick estimate of $\check A(\check t)$ using $\check A(\check t) \simeq \sqrt{2}\check u_{\rm rms}(\check t)$; this approximation would be exact if the mean flow were exactly sinusoidal (which is not the case in the GSF-dominated simulations) and if the perturbations were much smaller than the mean (which is not the case in the weakly stratified simulations). We checked that turbulent mixing does not affect the stratification significantly in the statistically stationary state of each simulation, so $N^2$ remains roughly constant in time and space.

 With these simplifications and caveats in mind, the trajectory of the fiducial simulation on the $(\Ric,\Ros^{-1})$ diagram is shown in Figure \ref{fig:plotshowcase}(d). We see that it starts at $\check t = 0$ on the top right inverted triangle when $\check A = 1$, which is in the linearly stable region, then moves toward the bottom left on a straight line. Note that because the only varying quantity in these expression is $\check A(t)$, simulation trajectories satisfy $\Ros_{\rm eff}^{-1} \propto \Ric_{\rm eff}^{1/2}$ at all times, and therefore appear as straight lines in the $(\Ric,\Ros^{-1})$ stability diagram (which uses logarithmic axes).
 When the line intersects the GSF-unstable region the flow becomes unstable, and then continues to move downward and to the left as the mean flow continues to increase slowly in amplitude (see above). The trajectory stops before it reaches the shear-unstable region (where the laminar steady state solution resides, and is marked by an upright triangle). As a result, and consistent with the findings above, the statistically stationary state is one that is dominated by the GSF instability. We note that the system could also be nonlinearly unstable to the diffusive shear instability at these parameters. However, visual inspection of the simulation shows that this does not appear to be the case (see also section \ref{sec:numquant}).

It is important to keep in mind that this visualization of the simulation trajectory should only be used for qualitative purposes. Indeed, the background image of $\log_{10} \check k_y$ on which the trajectory is superimposed was produced using a fixed Reynolds number $\Rey = 10000$ and a fixed P\'eclet number $\Pec = 10$, whereas in practice both $\Rey_{\rm eff}$ and $\Pec_{\rm eff}$ evolve with time with $\check A$. A more useful visualization can be created in the 3D parameter space of $(\Ric,\Ros^{-1},\Rey)$ (noting that $\Pec_{\rm eff} = \Pr \Rey_{\rm eff}$ where the Prandtl number is constant), but this is difficult to show in a printed figure.

In what follows, we now analyze a number of simulations with widely varying parameters, and attempt to characterize the dynamics observed based on the tools and arguments presented in this section.

\subsection{Exploration of parameter space}

We fix $\Rey_F = 100$ and $\Pec_F = 0.1$ (so $\Pra=0.001)$, and vary both $\Ric_F$ and $\Ros_F^{-1}$ to explore parameter space. 
Table \ref{tab:outs} lists the input parameters and summarizes salient results  for all available simulations. In all cases, the non-dimensional domain size is $(4\pi,2\pi,2\pi)$. Each simulation has a numerical resolution of $384\times192\times192$ equivalent grid points and is either initialized from small random perturbations to a laminar solution, or from the endpoint of another simulation ran at slightly different input parameters. 

\begin{table*}
  \singlespace
  \tabcolsep=0.11cm
  \centering
  \caption{Salient properties of all available simulations. Columns 1 and 2 list the governing parameters $\Ros_F^{-1}$ and $\Ric_F$ of each simulation. The remaining parameters $\Rey_F=10^2$ and $\Pec_F=0.1$ are the same in all cases. Columns 3 and 4 show the rms velocities $\check u_{\rm rms}$ and $\check w_{\rm rms}$ measured in the statistically steady state of each simulation (except the ones with $\Ric_F=4000$, see Table notes for detail). Column 5 shows the shearing rate $\check S_{\rm mid}$ extracted from the midplane of the domain (see Section \ref{sub:compturbvisc}). Column 6 shows the turbulent viscosity computed using $\check \nu_{\rm turb}=-(\check u\check w)_{\rm mid}/\check S_{\rm mid}$ (also see Section \ref{sub:compturbvisc}). }
\sisetup{separate-uncertainty=true}
\begin{tabular}{ccS[table-format=2.2(2)]S[table-format=1.3(3)]S[table-format=-2.1(1)]S[table-format=1.3(3)]}
\toprule
\toprule
\multicolumn{1}{l}{$\Ros_F^{-1}$} & \multicolumn{1}{l}{$\Ric_F$} & $\check u_{\rm rms}$ & $\check w_{\rm rms}$ & $\check S_{\rm mid}$ & $\check \nu_{\rm turb}$ \\
\midrule
0.2   & 1     & 1.8(2) & 1.19(9) & -1.7(4) & 0.7(2) \\
0.2   & 10    & 2.8(1) & 1(5)  & -2.7(6) & 0.5(1) \\
0.2   & 100   & 4.92(7) & 0.89(3) & -5.4(6) & 0.2(3) \\
0.2   & 1000  & 16.1(2) & 0.79(2) & -13.5(8) & 0.1(1) \\
0.2$^{\rm (a)}$   & 4000  & 34.4(4) & 0.061(2) & -37(7) & 0.014(5) \\
0.2$^{\rm (b)}$   & 4000  & 28.9(4) & 0.8(1) & -25(1) & 0.07(2) \\
0.2   & 10000 & 57.3(7) & 0.033(2) & -68(6) & 0.006(2) \\
1     & 1     & 1.9(1) & 1.4(1) & -1.6(4) & 0.7(3) \\
1     & 10    & 2.3(3) & 1.12(9) & -1.9(6) & 0.8(3) \\
1     & 100   & 4.9(2) & 0.96(6) & -4.3(6) & 0.34(7) \\
1     & 1000  & 13.55(8) & 0.82(2) & -11.4(8) & 0.13(1) \\
1     & 10000 & 34.5(2) & 0.43(4) & -33(1) & 0.03(4) \\
5     & 1     & 2(4)  & 2(4)  & -1.8(6) & 0.5(5) \\
5     & 10    & 4.1(9) & 1.5(4) & -4(1) & 0(3) \\
5     & 100   & 7.6(1) & 1.35(7) & -6(4) & 0.26(4) \\
5     & 1000  & 11.29(4) & 0.95(2) & -9(7) & 0.17(2) \\
5     & 10000 & 23.3(2) & 0.59(1) & -20.9(9) & 0.059(5) \\
5     & 100000 & 44.6(9) & 0.12(1) & -50(5) & 0.012(2) \\
\bottomrule
\end{tabular}%
  \label{tab:outs}%
  \begin{tablenotes}
        \item [] (a) data extracted in the time interval $(\check t-\check t_{\rm ref})\in[31.2,38.2]$ (see Fig. \ref{fig:plotrel-osc})
        \item [] (b) data extracted in the time interval $(\check t-\check t_{\rm ref})\in[44.2,51.2]$ (see Fig. \ref{fig:plotrel-osc})
      \end{tablenotes}
\end{table*}%

\begin{figure*}
     \caption{Effective parameters of a simulation plotted over the linear stability maps computed with $\Pec=10,\Rey=10000$ (left) and $\Pec_F=0.1,\Rey_F=100$ (right). The top row corresponds to the simulations discussed in Section \ref{sec:numsim}, while the bottom row shows other available simulations (see Table \ref{tab:outs}). The upward and inverted triangles denote the position in parameter space of the laminar solution $(\Ric_{\rm eff} = \Ric,\Ros^{-1}_{\rm eff} = \Ros^{-1})$ and the fully turbulent solution $(\Ric_{\rm eff} = \Ric_F,\Ros^{-1}_{\rm eff}= \Ros_F^{-1})$ for each simulation, respectively. The thin gray line connecting an upward triangle to the inverted triangle of the same color represents the maximum possible extent of a trajectory (see Section \ref{sec:charsim}). The thicker colored line overlaid on the gray line is the actual extent of the trajectory during the statistically steady state of the particular simulation. }
     \label{fig:maps}
     \centering
     \includegraphics[width=\textwidth]{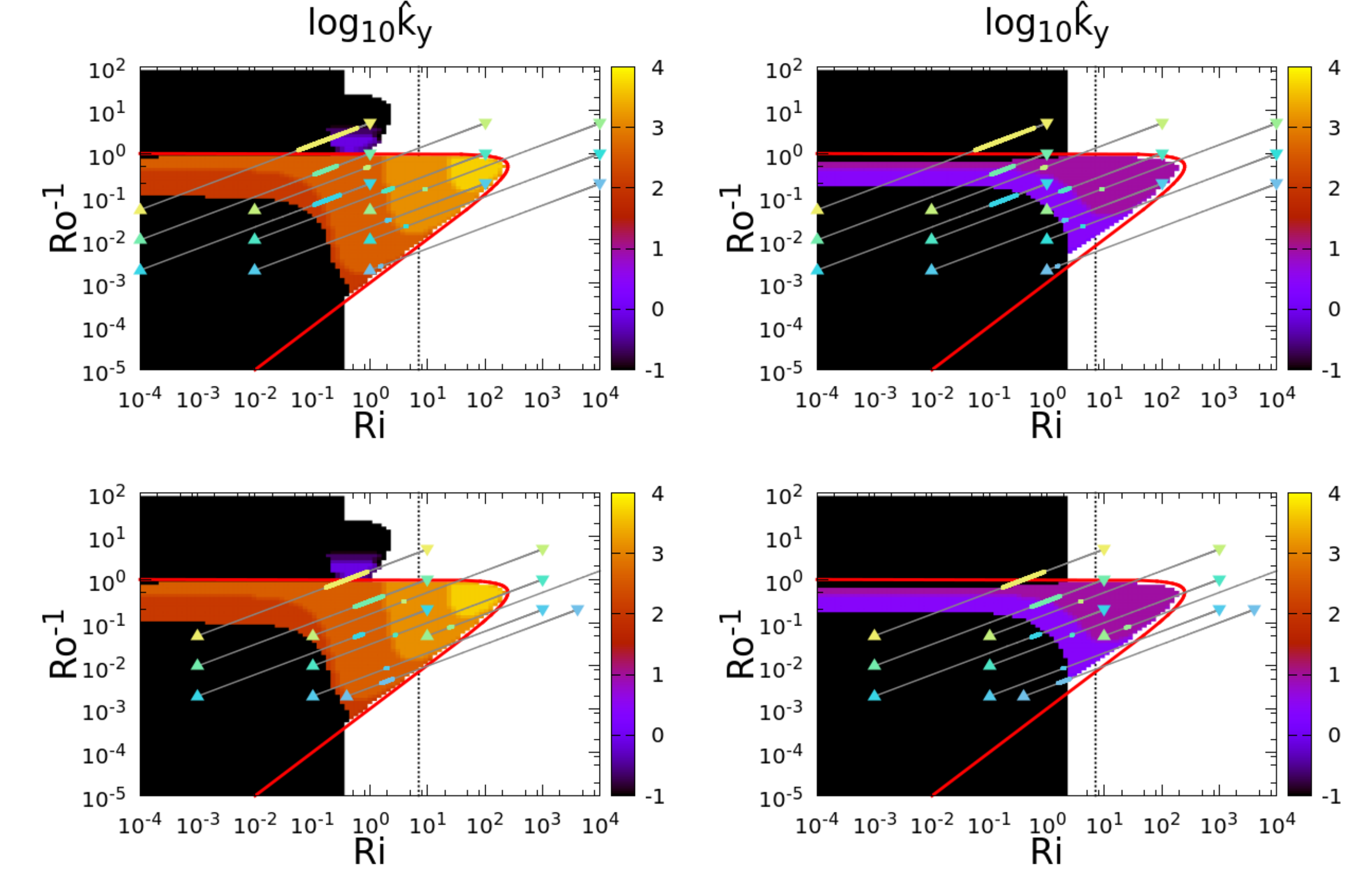}
\end{figure*}

To illustrate the wide range of possible emergent dynamics, we consider a subset of the data for three possible values of the input  Richardson number ($\Ric_F = 1, 100, 10000$), and three possible values of the input inverse Rossby number ($\Ros_F^{-1} = 0.2, 1, 5$). Figure \ref{fig:maps} (top row) shows where these simulations lie in parameter space as follows. Each simulation is assigned a color. The full possible extent of its trajectory in parameter space is shown as a thin gray segment of slope 1/2 (see the discussion of Figure \ref{fig:plotshowcase}(d)), ranging between the laminar state ($\check A = \Rey_F$, upright triangle) and the most turbulent state ($\check A = 1$, inverted triangle). For each  simulation, we waited until the system reached a statistically stationary state, then plotted the range of the trajectory in that state. With that choice, the simulation appears as a colored point if fluctuations in $\check A(\check t)$ are small, and as a colored segment if fluctuations in  $\check A(\check t)$ are large. The left column of Figure \ref{fig:maps} shows this information on a background color map of $\log_{10} \hat k_y$ obtained from a linear stability analysis using $\Rey_{\rm eff} = \Rey = 10000$, $\Pec_{\rm eff} = \Pec=10$, while the right column shows the same information superimposed on a similar color map obtained using $\Rey_{\rm eff} = \Rey_F = 100$, $\Pec_{\rm eff} = \Pec_F =0.1$. Comparing the two maps provides an idea of whether one can reliably identify a mode as being ``shear-dominated'' or ``GSF-dominated'' using linear stability analysis alone, showing that in some of the more clear-cut cases we can, but that in general we cannot (see more on this below). 
 
\begin{figure*}
     \caption{Representative snapshots of $\check u$ from simulation data for $\Rey_F=100,\Pec_F=0.1$ ($\Rey=10^4, \Pec=10$). From top to bottom: $\Ros_F^{-1}=5,1,0.2$ ($\Ros^{-1}=0.05,0.01,0.002$). From left to right: $\Ric_F=1,100,10000$ ($\Ric=0.0001,0.01,1$). In each snapshot, the long side of the domain corresponds to the streamwise direction ($x$), the vertical ($z$) is aligned with that of the page, and the remaining direction is that of the rotation axis ($y$).}
     \label{fig:simsnap0.1}
     \centering
     \includegraphics[width=0.7\textwidth]{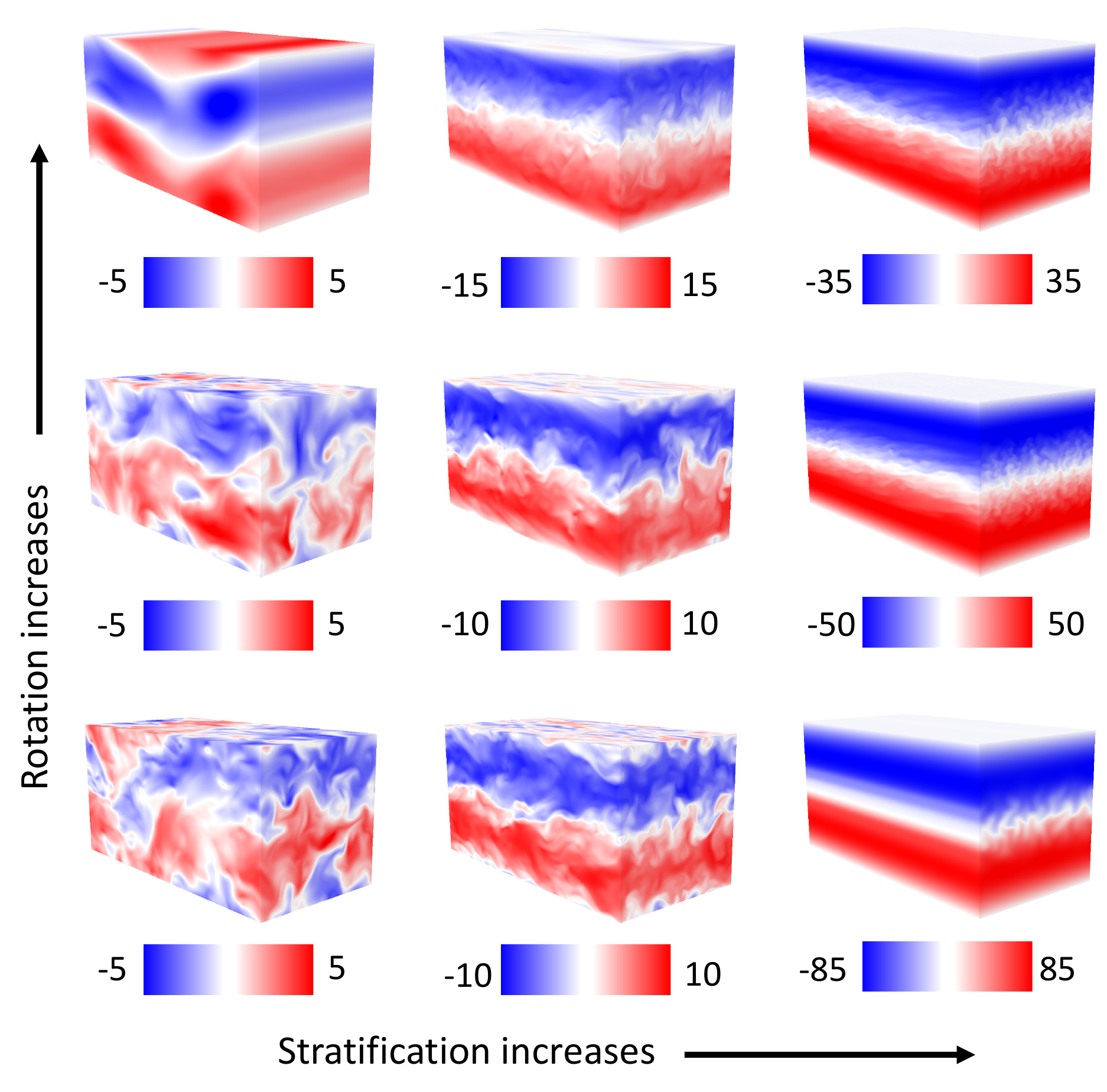}
\end{figure*}

\begin{figure*}
     \caption{Representative snapshots of $\check w$ from simulation data for $\Rey_F=100,\Pec_F=0.1$ ($\Rey=10^4, \Pec=10$). From top to bottom: $\Ros_F^{-1}=5,1,0.2$ ($\Ros^{-1}=0.05,0.01,0.002$). From left to right: $\Ric_F=1,100,10000$ ($\Ric=0.0001,0.01,1$). The orientation of the domain is the same as in Figure \ref{fig:simsnap0.1}.}
     \label{fig:simsnap0.1z}
     \centering
     \includegraphics[width=0.7\textwidth]{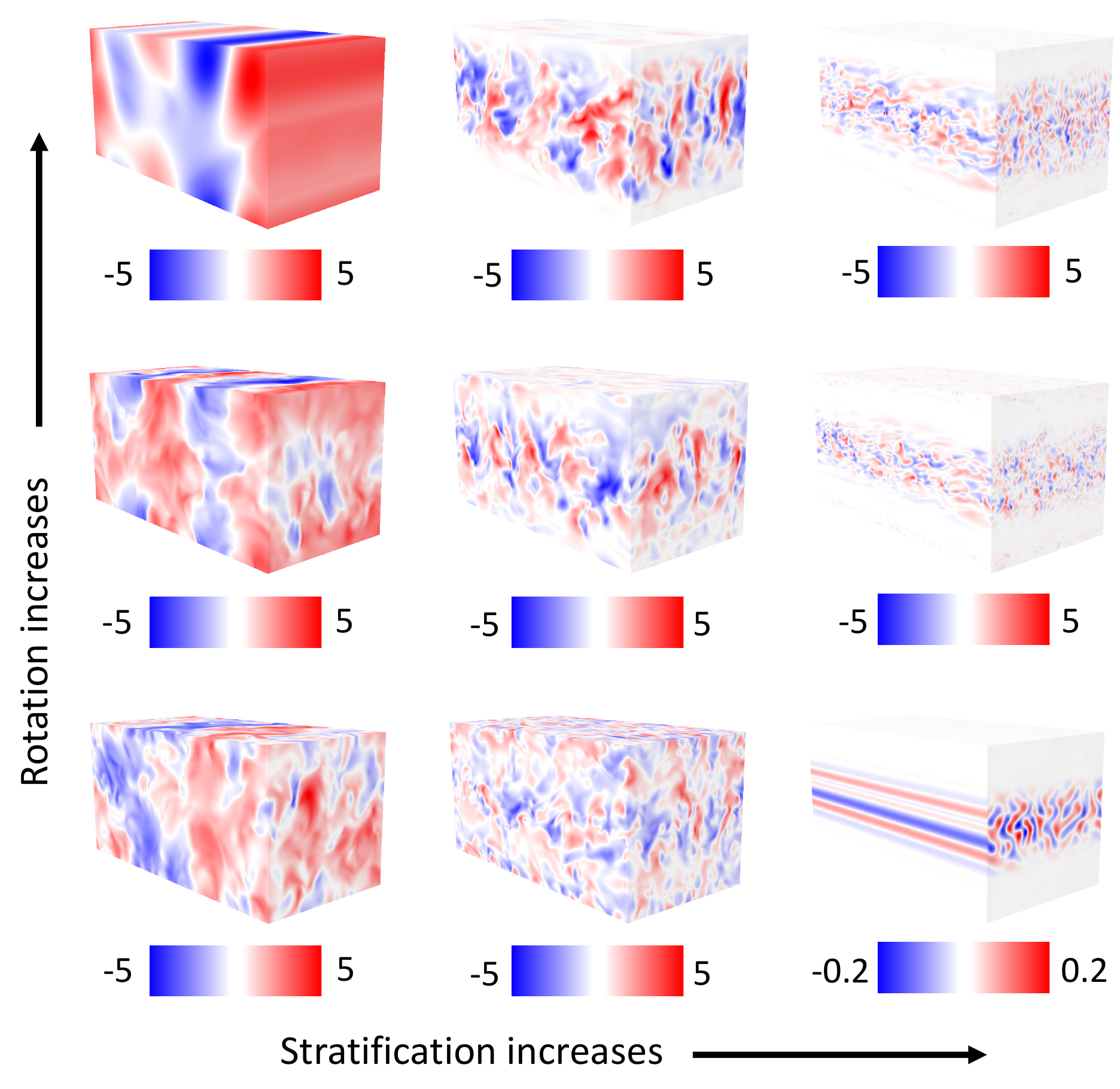}
\end{figure*}

Figures \ref{fig:simsnap0.1} and \ref{fig:simsnap0.1z} show representative snapshots of $\check u$ and $\check w$, respectively, in the same simulations, once they have achieved a statistically  stationary state. The snapshots are arranged in the same way as the triangles in the top row of Figure \ref{fig:maps}: each row from bottom to top corresponding to an increasing value of the rotation rate ($\Ros^{-1}_F = 0.2, 1, 5$), and each column from left to right corresponding to an increasing value of the stratification ($\Ric_F=  1, 100, 10000$).

We begin by looking at the most weakly rotating simulations (bottom row of Figures \ref{fig:simsnap0.1} and \ref{fig:simsnap0.1z}). In the left and center panels, the effect of rotation appears at a first glance to be negligible and the snapshots are qualitatively similar to those obtained by \cite{garaud_turbulent_2016} in the non-rotating case. As the Richardson number increases from $\Ric_F=1$ (left panel) to what \cite{garaud_turbulent_2016} refer to as the strongly stratified limit ($\Ric_F=100$, center panel), Figure \ref{fig:simsnap0.1z} shows that the scale of vertical velocity fluctuations decreases significantly.  As discussed by \cite{zahn_circulation_1992} \citep[see also][]{garaud_turbulent_2017}, the typical size of vertical eddies in non-rotating low P\'eclet number stratified turbulence is controlled by a combination of stratification and thermal diffusion, and would be proportional to $(\Ric_{\rm eff} \Pec_{\rm eff})^{-1/2}$ in the model setup and non-dimensionalization used here. This scaling is qualitatively consistent with the decrease in eddy size observed in the snapshots. Since a smaller eddy scale implies a decrease in the turbulent viscosity, this in turn results in a substantial increase of the amplitude of the mean flow,  which is clearly seen in Figure \ref{fig:simsnap0.1}.

This somewhat simplistic discussion is however called into question if we look at the corresponding trajectories of these two simulations in Figure \ref{fig:maps}. We see that they both lie close to the boundary between GSF-dominated and shear-dominated regimes, and in the case of the $\Ric_F = 100$ simulation (aquamarine color), both maps actually suggest that it lies in the GSF region of parameter space. An alternative explanation for the small eddy size observed in this simulation could then naively be that it is undergoing GSF-driven turbulence instead. But that interpretation is immediately invalidated by the presence of turbulence in regions of positive shear. In short, this example demonstrates the difficulty of identifying, simply from the position of the simulation on the linear stability maps, or from snapshots of $\check w$, whether the turbulence is shear-driven or GSF-driven. We will revisit this topic below in Section \ref{sec:numquant}, where we will finally understand this particular simulation as one that is dominated by the {\it nonlinear} branch of the shear instability.


Beyond a certain threshold in $\Ric_F$, the dynamics change dramatically and become governed by almost two-dimensional GSF modes when $\Ric_F=10000$ (bottom right panel in Figures \ref{fig:simsnap0.1} and \ref{fig:simsnap0.1z}). This result is relatively easy to understand given the position of the solution in the regime map (see the bottom right blue point in Figure \ref{fig:maps}, top row): at these parameters, the shear instability is not linearly excited, and the GSF instability is very close to the marginal stability boundary and therefore barely supercritical. This likely explains why the modes do not become fully turbulent and, as we shall demonstrate in Section \ref{sec:outliers}, why the nonlinear shear instability is not triggered.

\begin{figure*}
     \caption{Horizontally averaged profiles of the mean flow $\check {\bar u}(z,t)$ at selected instants in time, taken during the statistically stationary state, for $\Rey_F=100,\Pec_F=0.1$ ($\Rey=10^4, \Pec=10$). Instantaneous profiles are shown in gray, and their time average is shown in blue. From top to bottom: $\Ros_F^{-1}=5,1,0.2$ ($\Ros^{-1}=0.05,0.01,0.002$). From left to right: $\Ric_F=1,100,10000$ ($\Ric=0.0001,0.01,1$). }
     \label{fig:u}
     \centering
     \includegraphics[width=0.7\textwidth]{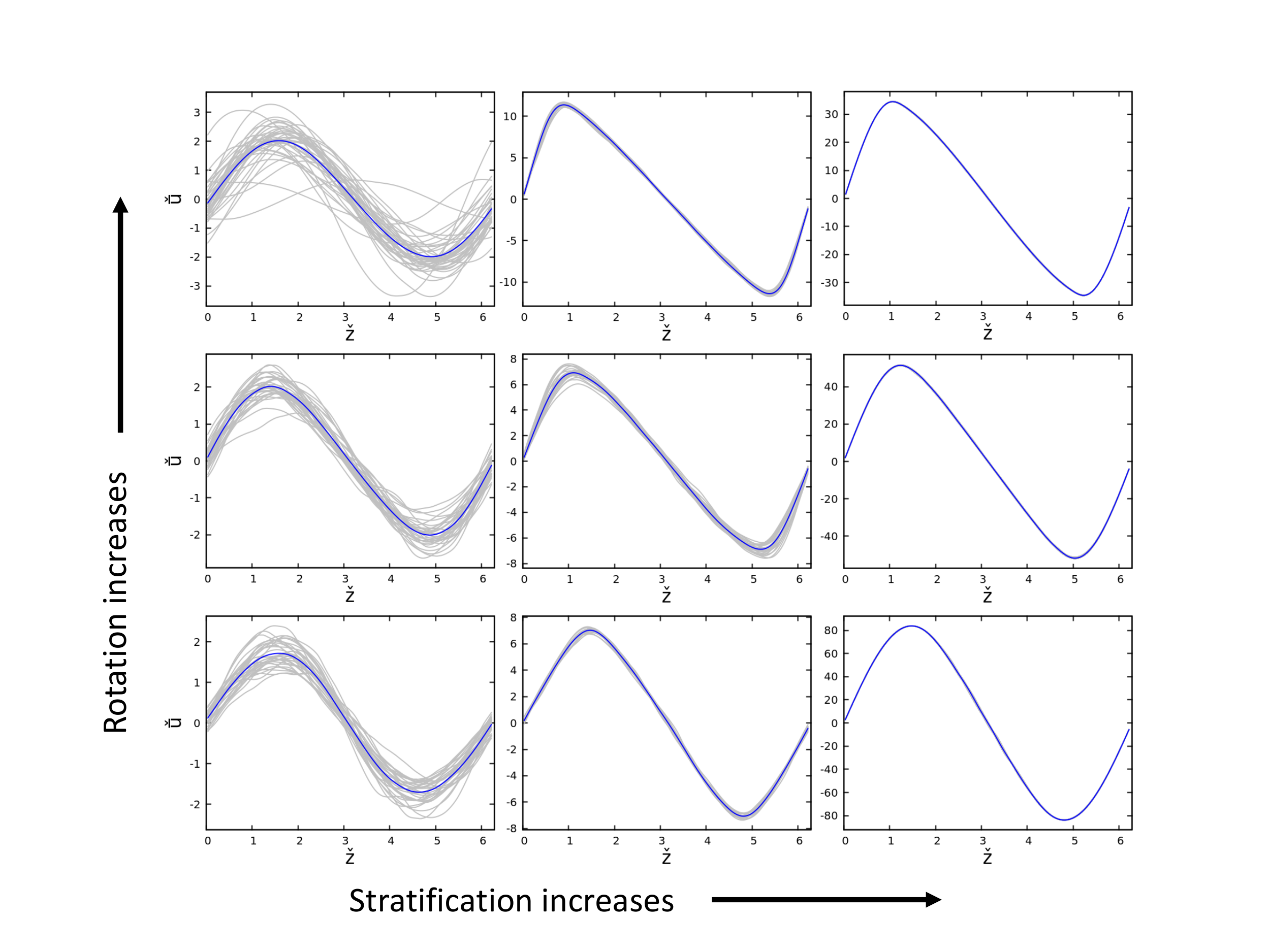}
\end{figure*}

A similar transition between shear-dominated dynamics at low Richardson number and GSF-dominated dynamics at higher Richardson number is observed when the rotation rate is a little higher (middle row in Figures \ref{fig:simsnap0.1} and  \ref{fig:simsnap0.1z}), i.e. when $\Ros_F^{-1}=1$. However, there are also a few important differences with the more weakly rotating case discussed above. In the leftmost panel of Figure \ref{fig:simsnap0.1z} (at $\Ric_F=1$), we see that the increased rotation rate causes the turbulence to be more coherent along the rotation axis. In the middle panel of the same row ($\Ric_F=100,\Ros_F^{-1}=1$), we see that the turbulence is becoming more inhomogeneous, somewhat suppressed in regions of positive shear, and enhanced in regions of negative shear (middle of the domain). In the rightmost panel of Figure \ref{fig:simsnap0.1z}, the system is now clearly dominated by the GSF instability. The turbulence has a very small vertical scale, and is limited to the region of negative shear. This strong inhomogeneity leads to a notable asymmetry in the mean flow, for the reasons discussed in Section \ref{sec:charsim}. This is illustrated in Figure \ref{fig:u} which shows, for each of the simulations of Figures \ref{fig:simsnap0.1} and  \ref{fig:simsnap0.1z}, a series of individual profiles of $\check {\bar u}(\check z,\check t)$, for various times $\check t$ selected once the system is in the statistically stationary phase. The layout is the same as in Figures \ref{fig:simsnap0.1} and  \ref{fig:simsnap0.1z}. We see that all simulations that are clearly dominated by the GSF instability have asymmetric mean flow profiles with a weaker negative shear in the turbulent region, and a stronger positive shear in the laminar regions, as in Figure \ref{fig:plotshowcase}(b). 

As we increase the rotation rate even further (top row in Figures \ref{fig:simsnap0.1}, \ref{fig:simsnap0.1z} and  \ref{fig:u}), the effect of rotation on the shear-dominated regime becomes quite pronounced. 
The more weakly stratified simulation ($\Ric_F=1$) now shows roll-like structures that are invariant in the $y$-direction. This reduction of the flow dynamics to two dimensions reverses the sign of the energy cascade, and the shear-induced rolls now span the entire vertical height of the domain as well. The mean flow profile in this regime is highly variable, as seen in Figure \ref{fig:u} (e.g. top left panel). However, this is only true at low $\Ric_F$, and as the stratification increases, the flow becomes 3D again and transitions to GSF-dominated dynamics.

To summarize our findings so far, at least from a qualitative point of view, we have found that the dynamics appear to be either shear-dominated or GSF-dominated depending on the parameters selected. This identification is sometimes trivial, as is the case for instance when the turbulence spans the entire domain (which would not happen in a GSF-dominated system), or in the case where the flow is invariant in the streamwise direction (which cannot extract any energy from the shear, and therefore could not be a shear instability). In other cases, by contrast, the identification can be much more difficult. We now turn to a quantitative analysis of momentum transport in our model system which, as we demonstrate below, provides a useful and more systematic way of characterizing the dynamics of a simulation.

\section{Quantitative analysis} 
\label{sec:numquant}

\subsection{Computing the turbulent viscosity}
\label{sub:compturbvisc}

\begin{figure*}
     \caption{Illustration of data extraction method. Instantaneous profiles of $d\check {\bar u}/d\check z$ and $\overline{\check u \check w}$, extracted during the statistically steady state, are shown in gray, and their time average is shown in blue. We fit constants to these quantities in the interval $[\pi-0.5,\pi+0.5]$ to measure $\check S_{\rm mid}$ and $(\check u\check w)_{\rm mid}$. The extracted means are shown in the solid red line, and the dotted red lines are placed one standard deviation above and below. The parameters for this simulation are: $\Ros_F^{-1}=1$ and $\Ric_F=100.$}
     \label{fig:measure_illust}
     \centering
     \includegraphics[width=0.8\textwidth]{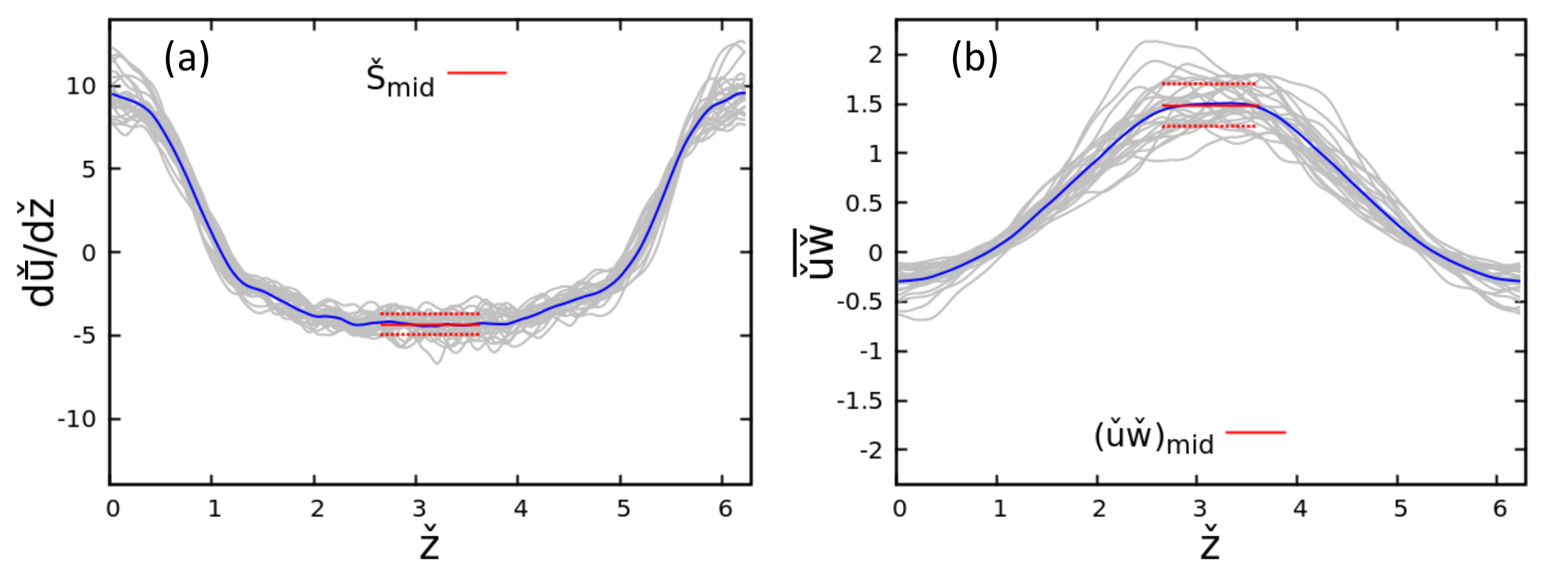}
\end{figure*}

 Assuming a linear relationship between the turbulent stress $\overline{\check u \check w}$ and the shearing rate $d\check {\bar u}/d\check z$ (see equation \ref{eqn:nu}), we can compute the turbulent viscosity $\check \nu_{\rm turb}$ by measuring $\overline{\check u \check w}$ and $d\check {\bar u}/d\check z$ from the DNS. We focus on the region near the middle of the domain where the shear is negative (so both instabilities can be present) and relatively constant. We therefore restrict our measurements to the interval $\check z \in [\pi-0.5,\pi+0.5]$. For each simulation,  we take all profiles of $d\check {\bar u}/d\check z$ that were saved during the statistically stationary state, and fit a constant to these profiles in the interval considered. The measured mean shearing rate is then denoted $\check S_{\rm mid}$ and the standard deviation around the mean yields an estimate of the measurement error and/or its variability. Similarly, we take all available profiles of $\overline{\check u \check w}$ during the statistically stationary state, and fit a constant to these profiles in the same interval. The measured mean stress is denoted $(\check u \check w)_{\rm mid}$, and its standard deviation is used to estimate the measurement error/variability. Both procedures are illustrated in Figure \ref{fig:measure_illust}, for the simulation with parameters $\Ros_F^{-1}=1$ and $\Ric_F=100$. From these measurements, we deduce the non-dimensional value of the turbulent viscosity in the middle of the domain, as 
\begin{equation}
\check \nu_{\rm turb} = - \frac{ (\check u \check w)_{\rm mid} }{\check S_{\rm mid}}.
\end{equation}

We can then compare our simulation data to the predictions of existing models, specifically those of \cite{garaud_turbulent_2017} for the turbulent viscosity of non-rotating low P\'eclet number shear flows, and of \cite{barker_angular_2019} for the turbulent viscosity induced by the GSF instability. 


\subsection{Comparison with the shear instability model of Garaud et al. (2017)}
\label{sec:shearcomp}


The model of  \cite{garaud_turbulent_2017} provides the following estimate\footnote{On further inspection, we discovered that the formula given in equation (42) of \cite{garaud_turbulent_2017} is actually not quite correct, using $(J\Pec)^{-1}$ in the denominator of the first term instead of $(J\Pec_F)^{-1}$. The formula provided here corrects this error. The constant $a$ needed to be re-fitted accordingly.} for the dimensional turbulent viscosity $\nu_{\rm turb}$ in non-rotating, stratified shear flows at low input P\'eclet number $\Pec_F$:
\begin{gather}
    \alignedb
    \nu_{\rm turb} = \frac{C}{1+a(J\Pec_F)^{-1}} \left( 1 - \frac{J\Pra}{(J\Pra)_c} \right)^b \frac{\kappa_T}{J} \quad {\rm for}\ J\Pra < (J\Pra)_c, 
    \alignede
    \label{eq:nuturbgagnier}
\end{gather}
where $J$ is the local Richardson number (which here is equal to $\Ric_F / \check S_{\rm mid}^2$), $\Pra$ is the Prandtl number (which is equal to $0.001$ in our DNS), and where  
$C\approx a \approx 0.08, \ b\approx0.25,$ and $(J\Pra)_c\approx0.007$ are model constants that were fitted to the non-rotating data  \citep[see][]{garaud_turbulent_2017}. Non-dimensionally, this becomes 
\begin{gather}
    \alignedb
     \check \nu_{\rm turb}  = \frac{C}{a+J\Pec_F} \left( 1 - \frac{J\Pra}{(J\Pra)_c} \right)^b  \quad {\rm for}\ J\Pra < (J\Pra)_c, 
    \alignede
    \label{eq:nuturbpred1}
\end{gather}
which tends to $C/a \simeq 1$ as $J \rightarrow 0$. This is expected from the non-dimensionalization selected, which assumes a balance between the turbulent stresses and the forcing. 

\begin{figure*}
     \caption{Comparison of $\check \nu_{\rm turb}$ and $(\check u \check w)_{\rm mid}$ (symbols) with the theoretical models (magenta lines) of \protect\cite{garaud_turbulent_2017} (left) and \protect\cite{barker_angular_2019} (right). The symbol color represents the rotation rate (see legend) and the symbol size is inversely related to the stratification. The gray point are the non-rotating data of \protect\cite{garaud_turbulent_2016}. The square blue points correspond to a particular simulation at $\Ric_F = 4000$ discussed in Section \ref{sec:priming}. Note the quantity plotted on the right panel, $-(\check u \check w)_{\rm mid}/(\Ros_F^{-1}+\check S_{\rm mid})$ is essentially the same as $\check \nu_{\rm turb}$ when $\Ros_F^{-1}\ll \check S_{\rm mid}$.}
     \label{fig:plotmodels}
     \centering
     \includegraphics[width=\textwidth]{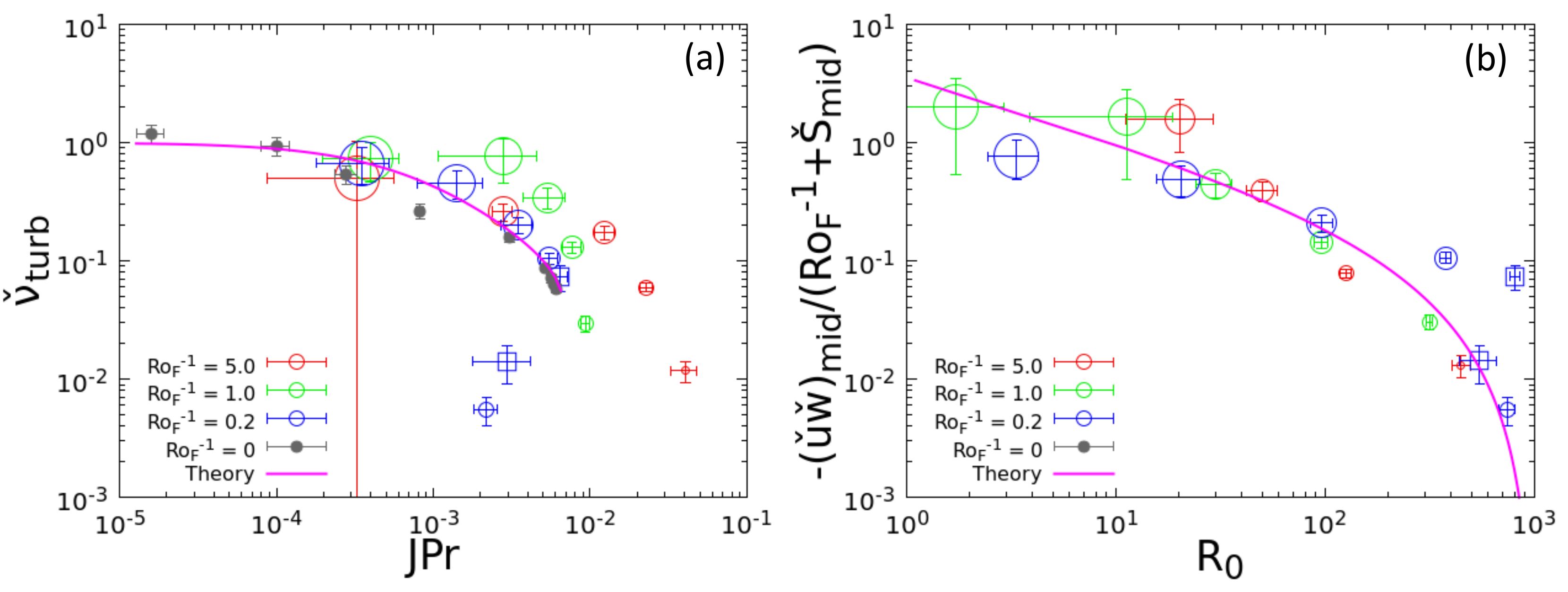}
\end{figure*}

Figure \ref{fig:plotmodels}(a) shows the quantity $\check \nu_{\rm turb}$ computed from the DNS in the manner described in the previous section, as a function of $J\Pra$. It is compared with the model prediction (magenta curve, which represents the right-hand side of equation \ref{eq:nuturbpred1}). Note that since $J \Pec_F = J\Pra \Rey_F$, with $\Rey_F = 100$ being held constant, the right-hand side of this equation is a function of $J \Pra$ only in our dataset, hence our choice to present the results as a function of this quantity. Plotted in this manner, all simulations dominated by the shear instability should lie on (or close to) the magenta curve. The non-rotating DNS from \cite{garaud_turbulent_2016}, which were run at the same values of $\Rey_F$ and $\Pec_F$ as our rotating simulations, are shown as small solid gray circles. We see that the model fits these non-rotating simulations well, as expected, except for a little dip near $J\Pra \sim 0.001$, which seems to be a real feature of the data \citep[see][]{garaud_turbulent_2017}. Simulations for increasing rotation rates are shown as colored circles (see legend for detail), and the size of the circle is linearly related to $-\log_{10}(\Ric_F)$ (so the largest circles correspond to the least stratified simulations at $\Ric_F = 1$, and the smallest circles correspond to the most strongly stratified simulations). The data reveals very interesting, if sometimes puzzling trends. 

\subsection{Slowly rotating diffusive shear flows}

As expected, most of the slowly rotating simulations ($\Ros_F^{-1} = 0.2$, blue circles) lie close to the non-rotating data and the magenta model curve. We also see a few outliers, which have $\Ric_F = 4000$ and $10000$, respectively, that turn out to be particularly interesting (see Section \ref{sec:outliers} below). Ignoring these outliers for now, our results tentatively confirm that turbulent transport for $\Ros_F^{-1} = 0.2$ (and lower) is primarily shear-driven, and is consistent with the fact that the turbulence spans most of the computational domain (including, crucially, regions where the shear is positive) in the corresponding snapshots of Figure \ref{fig:simsnap0.1z}. 

However, it is important to note that, with the exception of the least stratified case ($\Ric_F = 1$), the mean flow is {\it linearly stable} to the shear in these simulations. As such, the dynamics observed in simulations for $\Ric_F = 10, 100$, and $1000$ must be driven by {\it nonlinear} shear instabilities. It is easy to verify that our findings are  consistent with Zahn's instability criterion (see \cite{zahn_rotational_1974}, \cite{garaud_turbulent_2017} and Section \ref{sec:intro}): indeed, the blue points all lie in the region where $J\Pra < (J\Pra)_c \simeq 0.007$. Interestingly, some of them get very close to this stability threshold, suggesting that weak rotation does not affect it.

In summary, we find that for weak rotation, and moderate stratification, existing models for turbulent mixing by diffusive stratified shear flows hold. The same is not true, however, for larger rotation rates and/or very strong stratification.

\subsection{Comparison with the GSF instability model of Barker et al. (2019)}
\label{sec:barkercomp}

For more rapidly rotating simulations (higher $\Ros_F^{-1}$), we see that, with a few exceptions at lower values of the stratification, the data does not fall on the model curve for shear-induced turbulence, and instead lies above it (most red and green points). This is not  surprising since many of these simulations were tentatively identified as being GSF-dominated in the previous section. We do note that the GSF instability not only persists for $J\Pra > 0.007$ (which is not unreasonable, since it is not limited by Zahn's criterion), but can also have a turbulent viscosity that is sometimes substantially larger than that of the pure shear instability. To see why this is the case, we now  compare our data with the GSF model of  \cite{barker_angular_2019}. 

\cite{barker_angular_2019} studied turbulent transport by the GSF instability, and proposed a model for the turbulent viscosity at saturation that relies on balancing the growth rates of primary instability and parasitic instabilities \citep[see also][for related work on the analogous fingering instability]{brown_chemical_2013}. Dimensionally, when written using the notation of this paper, their model predicts that the Reynolds stress should be
\begin{equation}
    \overline{ u w} = - C_{B}^2 \frac{S_{\rm mid} + 2 \Omega}{ \lambda  + \nu k_y^2 } \frac{\lambda^2}{k_y^2}, 
\end{equation}
where $\lambda$ and $k_y$ are the dimensional growth rate and wavenumber of the fastest-growing GSF mode, $S_{\rm mid}$ is the dimensional value of the shearing rate (taken here in the middle of our domain, where it is negative), and $C_{B}$ is a constant of order unity that needs to be fitted to the data. \cite{barker_angular_2019} find\footnote{ More specifically, they find that a best fit to their data is obtained with $A\simeq 4$ (see their equation 31), which corresponds to $C_B \simeq \sqrt{8}$ using the relation $A^2/2=C_B^2$.} that $C_B\simeq \sqrt{8}$. We can express this in our code units to get a prediction for the Reynolds stress in the middle of the domain:  
\begin{equation}
    ( \check u \check w)_{\rm mid} = - C_{B}^2 (\check S_{\rm mid} + \Ros_F^{-1}) \frac{\check \lambda^2}{\check k_y^2 (\check \lambda  + \Rey_F^{-1} \check k_y^2 )} .
    \label{eq:gsfbarker}
\end{equation}
We know from the analogy between the GSF and the fingering instabilities, that when $\lambda$ and $k_y$ are written in the natural units for double-diffusive convection (namely, $d = (\kappa_T \nu / N^2)^{1/4}$ as the unit length, and $d^2/\kappa_T$ as the unit time), they are only 
functions of the Prandtl number, and of the so-called equivalent density ratio, defined in our units as 
 \begin{gather}
    R_0=-\frac{\Ric_F}{\Ros_F^{-1}(\Ros_F^{-1}+\check S_{\rm mid})}
    \label{eq:R0nondim}
\end{gather}
(see Section \ref{sec:intro}). An alternative way of writing (\ref{eq:gsfbarker}) is therefore:
\begin{equation}
    -\frac{( \check u \check w)_{\rm mid}}{\Ros_F^{-1}+\check S_{\rm mid}} = \frac{C_{B}^2}{\Pec_F} \frac{ \lambda_{gsf}^2}{ k_{gsf}^2 ( \lambda_{gsf}  + \Pra k_{gsf}^2 )}, 
    \label{eq:barker_ourway}
\end{equation}
where $\lambda_{gsf}$ and $k_{gsf}$ are the growth rate and wavenumber of the fastest-growing GSF modes  expressed in their natural units (so $\lambda_{gsf} = \lambda d^2/ \kappa_T $ and $k_{gsf} = d k_y$). 
Using this expression has two advantages. First, note that the right-hand-side of (\ref{eq:barker_ourway}) is a function of $R_0$ only when the Prandtl number is fixed, which is the case of our simulations. As such, and as demonstrated by \cite{barker_angular_2019}, plotting the quantity $-(\check u \check w)_{\rm mid}/(\Ros_F^{-1}+\check S_{\rm mid})$ against $R_0$ should collapse the data on a single curve if the system is only subject to the GSF instability. Second, note that  $-(\check u \check w)_{\rm mid}/(\Ros_F^{-1}+\check S_{\rm mid}) \simeq -(\check u \check w)_{\rm mid}/\check S_{\rm mid} \simeq \check \nu_{\rm turb}$ when $\Ros_F^{-1} \ll |\check S_{\rm mid}|$, which is often the case in our simulations. As a result, plotting $-(\check u \check w)_{\rm mid}/(\Ros_F^{-1}+\check S_{\rm mid})$ is almost the same as plotting $\check \nu_{\rm turb}$, which allows for an easy comparison with Figure \ref{fig:plotmodels}(a).

Figure \ref{fig:plotmodels}(b) shows $-(\check u \check w)_{\rm mid}/(\Ros_F^{-1}+\check S_{\rm mid})$ against $R_0$ computed using (\ref{eq:R0nondim}). The symbols used for each simulation are identical to the ones in Figure \ref{fig:plotmodels}(a), with the color representing $\Ros_F^{-1}$ and the size representing $\Ric_F$.
The solid magenta line is the model prediction from equation (\ref{eq:barker_ourway}), with the constant fitted to the data equal to $C_B\simeq\sqrt{8}$. Figure \ref{fig:plotmodels}(b) reveals a number of interesting features of our data. 

First, note that not all simulations depicted in Figure \ref{fig:plotmodels}(a) are present in \ref{fig:plotmodels}(b) -- the largest red point, corresponding to a run with $\Ros_F^{-1} = 5$, $\Ric_F=1$, is missing. This is because in this case, $R_0$ is smaller than one, and the system is not subject to GSF instabilities (consistent with the position of the simulation in the stability diagrams, see Figure \ref{fig:maps}).  

Second, looking at the remaining data points present in Figure \ref{fig:plotmodels}(b), we see that the turbulent viscosity model for the GSF instability  provides a good explanation for the data in many, but crucially not all cases. In particular, we see a few blue points (for $\Ros_F^{-1} = 0.2$) that clearly lie above the GSF model curve at large $R_0$. These points correspond to those that are in the shear-unstable branch very close to $J\Pra = 0.007$ in Figure \ref{fig:plotmodels}(a), with $\check \nu_{\rm turb} \simeq 0.1$. We therefore confirm that these simulations are dominated by the shear instability, and that the turbulent viscosity in that case is much larger than that of the GSF. Conversely, we now also see that the blue outliers that were lying well-below the model curve for the shear instability in Figure \ref{fig:plotmodels}(a), are well-explained by the GSF model curve in \ref{fig:plotmodels}(b) (these are the blue points at very large $R_0$ that lie almost on top of the GSF model curve). 

Finally, we were initially surprised to see that the GSF model curve fits the data for small $R_0$ quite well in Figure \ref{fig:plotmodels}(b), even when a simulation was identified to be in the shear-dominated regime. After further investigation, we discovered that this is most likely a coincidence that accidentally arose from our choice of parameters. Indeed, in the limit $R_0 \rightarrow 1$, one can use the asymptotic scalings derived by \cite{brown_chemical_2013} to show\footnote{See equation B5 of \cite{brown_chemical_2013} using $\phi = 1$, because the equivalent of the diffusivity ratio $\tau$ is the Prandtl number for GSF modes.} that $\lambda_{gsf} \simeq \sqrt{\Pra}$ and $k_{gsf} \simeq 1/\sqrt{2}$. With this, (\ref{eq:barker_ourway}) implies that 
\begin{equation}
    -\frac{( \check u \check w)_{\rm mid}}{\Ros_F^{-1}+\check S_{\rm mid}} \simeq \check \nu_{\rm turb} \simeq 2 \frac{C_{B}^2}{\Pec_F} \sqrt{ \Pra} \mbox{ when } R_0 \rightarrow 1, \Pra \ll 1,
\end{equation}
(which is the case for our simulations, since $\Pra = 0.001$). With $C_{B}^2 \simeq 8$, and $\Pec_F= 0.1$, this predicts that $\check \nu_{\rm turb} \simeq 5$ in weakly stratified systems which is quite close to what the shear instability model predicts in the same limit, but only coincidentally. Had we selected substantially different values of $\Pec_F$ or $\Pra$, the GSF predictions and shear predictions would have been quite different, and we believe this would be more clearly visible in the data. As it is, we do see that the data at low $R_0$ is more consistent with being almost constant (which the shear model predicts) than with the GSF model, but this will need to be verified in the future with a more comprehensive exploration of parameter space.

\subsection{Summary so far}

To summarize our results, we find that the model of \cite{barker_angular_2019} correctly predicts the turbulent viscosity  measured in a simulation whenever it is  dominated by the GSF instability. Similarly, the model of \cite{garaud_turbulent_2017} correctly predicts the measured turbulent viscosity whenever a simulation is dominated by the shear instability. Taken on its own, this result is superficially pleasing but does not answer the more important question of {\it when} or {\it why} a simulation ends up being dominated by one instability or the other when both can theoretically be excited. A very naive approach would be to compare the linear growth rates of each mode of instability and select whichever is largest, but this would obviously not work here -- in strongly stratified flows, the shear instability is primarily excited through a nonlinear pathway, that cannot be captured in this manner. Ignoring the outliers (see below for more on these points), an alternative empirical answer to this question may be the following: the instability that ends up dominating is the one that would individually contribute the most to turbulent transport. In other words, one could compute the turbulent viscosity predicted by the GSF model of \cite{barker_angular_2019}, as well as the one predicted by the shear instability model of \cite{garaud_turbulent_2017}, and whichever one is the largest identifies the dominant instability. This method, when applied to our data, would correctly identify almost all simulations (and therefore also correctly predict the measured turbulent viscosity), {\it except for the outliers}. 
It is therefore time to take a closer look at these simulations, to see what is happening in this case.

\subsection{The outliers}
\label{sec:outliers}

The blue square and blue circle, found substantially below the magenta curve in Figure \ref{fig:plotmodels}(a), are from very strongly stratified simulations with $\Ric_F = 4000$ and $\Ric_F = 10000$, respectively. 
These simulations are clearly GSF-dominated, despite satisfying Zahn's instability criterion ($J\Pra<0.007$), and despite the fact that at these parameters the model of \cite{garaud_turbulent_2017} would predict a much larger turbulent viscosity than the model of \cite{barker_angular_2019}. This can be seen either from the individual snapshot in Figure \ref{fig:simsnap0.1z}, which reveals the flow to be that of a two-dimensional mode of the GSF instability, or from the corresponding position of the data points on Figure \ref{fig:plotmodels}(b), where they lie almost exactly on top of the GSF model curve. The existence of these points therefore appears to  directly contradict the proposal made above to identify which instability ought to dominate.

More worryingly, they also show that it is possible to have two stratified rotating shear flows with similar values of $J\Pra$ and the same rotation rate, but with two very different values of the turbulent viscosity -- in the cases shown here, $\check \nu_{\rm turb}$ in the GSF-dominated simulations is up to two orders of magnitude smaller than in the shear-dominated simulations at the same value of $J\Pra$ and $\Ros_F^{-1}$. 
Of course, it is important to remember that merely satisfying Zahn's instability criterion does not guarantee that a system will be subject to shear-induced turbulence. Since the instability is subcritical in that regime, its development relies on the availability of finite-amplitude perturbations of the right kind and of sufficient amplitude to ``prime'' the turbulence. In the non-rotating case, as discussed in the introduction, this has been shown to lead to the existence of hysteresis in the system \citep{garaud_stability_2015,garaud_turbulent_2016,gagnier_turbulent_2018}, with otherwise similar simulations being turbulent or not depending on the manner in which they were initialized. It is therefore natural to find that the same phenomenon occurs in the rotating case, and the failure to trigger shear-induced turbulence in these outlying blue points is likely simply be due to the lack of a proper ``primer''. This realization brings us to the more interesting question of how to prime the nonlinear shear instability.

\subsection{Priming the shear instability}
\label{sec:priming}

In the non-rotating case, priming the nonlinear shear instability is quite difficult. \cite{garaud_turbulent_2016} and \cite{garaud_turbulent_2017} were only able to do it by  using as initial conditions the turbulent state of the system at a slightly lower stratification. In other words, the shear instability can persist into the nonlinear regime if the stratification increases very gradually, but disappears otherwise. Interestingly, we are finding that the situation is quite different in the rotating case, because the GSF instability can serve as a primer for the shear instability, as long as it is not two-dimensional. 

\begin{figure}
     \caption{Illustration of the GSF instability's ability to  prime the nonlinear shear instability. The quantities $\check u_{\rm rms}$ (left axis, top curves) and $\check w_{\rm rms}$ (right axis, bottom curves) are shown for two simulations with $\Ros_F^{-1}=0$ (blue and pink curves) and $0.2$ (red and green curves). At $\check t = \check t_{\rm ref}$, the stratification increases from $\Ric_F=1000$ to $\Ric_F=1700$.} 
     \label{fig:plotpriming}
     \centering
     \includegraphics[width=0.45\textwidth]{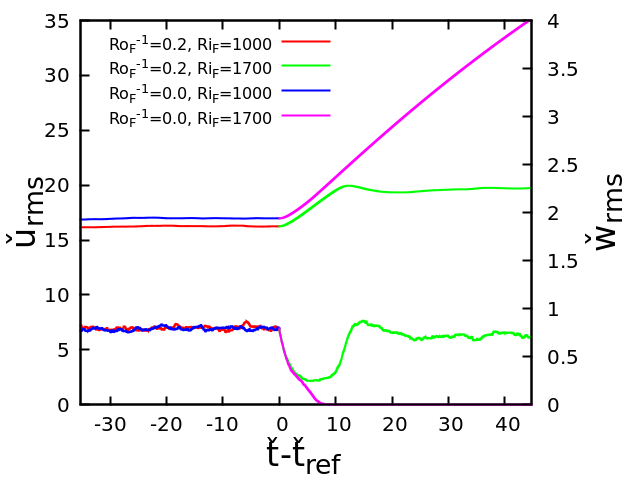}
\end{figure}

A simple demonstration of this effect is shown in Figure \ref{fig:plotpriming},  which summarizes a simple experiment in which the Richardson number is  suddenly increased at $\check t = \check t_{\rm ref}$ from $\Ric_F = 1000$ to $\Ric_F = 1700$, in two simulations that had reached a statistically stationary shear-dominated state, one non-rotating (blue and magenta curves), and one at $\Ros_F^{-1} = 0.2$ (red and green curves). We see that the turbulence, characterized for example by $\check w_{\rm rms}$, dies out in the non-rotating case, but rapidly recovers in the rotating case, and remains at a level consistent with that of shear-induced turbulence. Since the rotating case is GSF unstable, we conclude that the small-scale turbulence associated with the GSF instability must be able to prime the shear instability. In hindsight, this is not surprising. The GSF instability only exists because it is thermally diffusive (see Section \ref{sec:intro}), and therefore has a typical lengthscale that is small enough for thermal diffusion to take place.  This is precisely the characteristic eddy scale that is required to trigger the nonlinear shear instability, and we see in this example that it does.

When the background stratification continues to increase, however, the system eventually approaches the  marginal stability threshold for the GSF instability ($R_0 \rightarrow \Pra^{-1}$, see equation \ref{eq:instabscrit}). When this happens (e.g. for the simulation at $\Ric_F = 10000$, see Figure \ref{fig:maps}), the saturated ``turbulent'' state of the GSF remains two-dimensional, and is invariant in the streamwise direction. Since it is not possible to extract energy from the shear using streamwise-invariant perturbations, this two-dimensional form of the GSF instability cannot prime the shear instability, and the system remains in a GSF-dominated state. This explains the existence of the outliers discussed in Section \ref{sec:outliers}, and why these are only found for very large $R_0$.

\begin{figure}
     \caption{Illustration of relaxation oscillation dynamics that occur in a simulation at $\Ros_F^{-1}=0.2$ and $\Ric_F=4000$ (see text for detail), showing $\check u_{\rm rms}$ (left axis, red curve) and $\check w_{\rm rms}$ (right axis, green curve). The gray regions mark the two distinct time intervals over which the turbulent viscosity is measured, and shown as blue squares in Figure \ref{fig:plotmodels}.} 
     \label{fig:plotrel-osc}
     \centering
     \includegraphics[width=0.45\textwidth]{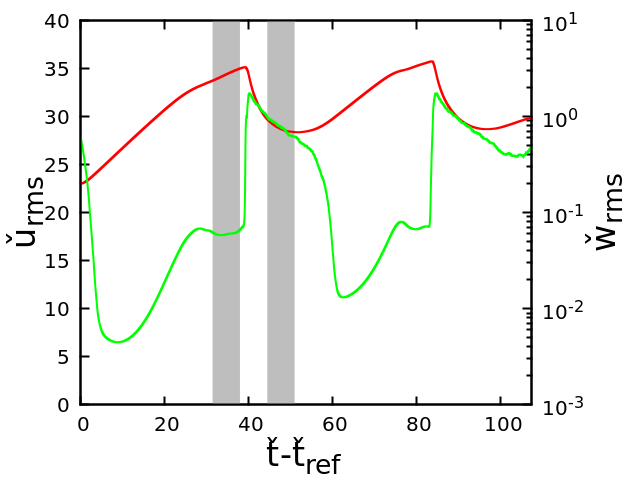}
\end{figure}

Finally, we also found that the interaction of the shear instability and the GSF instability through priming can drive relaxation oscillations, a result that was fairly unexpected. These oscillations are illustrated in Figure \ref{fig:plotrel-osc}, which shows both $\check u_{\rm rms}$ and $\check w_{\rm rms}$ as a function of time, in a simulation at $\Ros_F^{-1} = 0.2$, $\Ric_F = 4000$. The horizontal axis shows $\check t- \check t_{\rm ref}$, where $\check t_{\rm ref}$ in this case was arbitrarily selected to be the time origin once the system has entered this quasi-periodic state. The corresponding trajectory of this simulation on the stability map is shown in Figure \ref{fig:maps} (bottom row, bottom right blue/gray trace in each panel). The sequence of events associated with a single cycle of the oscillation (e.g. from $\check t - \check t_{\rm ref} \simeq 10$ to $60$ in Figure \ref{fig:plotrel-osc}) is as follows. At the start of the cycle, the system is in a state of weak shear, which is very close to being marginally stable to the GSF instability (near the red curve in Figure \ref{fig:maps}). This state is unstable to a slowly-growing,  two-dimensional GSF mode which eventually saturates (around $\check t - \check t_{\rm ref} \simeq 25$). 
A this point, the flow looks like that of the bottom right panel of Figures \ref{fig:simsnap0.1} and \ref{fig:simsnap0.1z}. The system remains in that GSF-dominated state, but the background shear continues to increase in response to the forcing (albeit more slowly now). When the shear is large enough, the GSF flow becomes three-dimensional, at which point the shear instability can finally be nonlinearly excited (around $\check t - \check t_{\rm ref} \simeq 40$). When this happens, the flow looks like the bottom middle panel of Figures \ref{fig:simsnap0.1} and \ref{fig:simsnap0.1z}. The turbulent viscosity increases dramatically, and the shear decreases suddenly as a result. This  happens too fast for the nonlinear shear instability to keep up (even though $J\Pra$ remains smaller than 0.007), and the latter dies down. The system moves back into the almost-marginally-stable two-dimensional GSF state, and the cycle repeats. This regular oscillation between a GSF-dominated state and a shear-dominated state can also be seen in Figure \ref{fig:plotmodels}. The square blue symbols both correspond to the same $\Ric_F = 4000$ simulation discussed here, but were extracted during distinct time intervals, marked in gray in Figure \ref{fig:plotrel-osc}. We see that while the flow is shear-dominated the turbulent viscosity is relatively high and satisfies the \cite{garaud_turbulent_2017} model, and when the flow is GSF-dominated the turbulent viscosity is low and satisfies the \cite{barker_angular_2019} model.

This is an interesting example of a relaxation oscillation driven by the nonlinear priming of one subcritical instability by another supercritical one. 
This oscillation between two turbulent states is accompanied by a significant oscillation in the mean flow amplitude. Assuming that these oscillations persist at lower Prandtl number, this mechanism could possibly be at the origin of shear oscillations in some stars that are close to the marginal stability threshold for the GSF instability.

\section{Summary and discussion}
\label{sec:con}

In this work, we studied stably stratified, thermally diffusive, {\it rotating} shear flows in the equatorial region of stars, building on previous work that had focused on the non-rotating case  \citep{prat_turbulent_2013,garaud_stability_2015,garaud_turbulent_2016,garaud_turbulent_2017,gagnier_turbulent_2018}. This extension, as introduced in Section \ref{sec:intro}, is necessary since the main source of large-scale shear in stars is their differential rotation. 

We used a very simple model setup, in which a body force drives a vertically varying azimuthal flow (see Section \ref{sec:model}). 
 A linear stability analysis of this model (see Section \ref{sec:linstabres})  reveals the existence of two modes of instability: shearing modes, that only depend on the velocity gradient, and GSF modes, that only depend on the angular momentum gradient. Interestingly, both kinds of modes are found to coexist in a substantial region of parameter space in our model, which naturally leads to the question of which instability dominates when both are excited. To complicate the matter further,  diffusive shear instabilities are known to be nonlinearly unstable (even when they are linearly stable), which means that they can coexist with the GSF instability in a much wider region of parameter space than what linear theory alone suggests.

Using DNS, we then studied the nonlinear development of these instabilities, and were able to measure the turbulent viscosity they cause for a wide range of input parameters (varying both the rotation rate and the stratification).  The limit of very low stratification, where the non-diffusive shear instability takes place, was briefly discussed for completeness, but is not particularly relevant for stellar interiors except perhaps very close to  the edge of a convection zone where the buoyancy frequency drops to zero. 

Much more relevant for stellar radiative zones is the limit of very large stratification (i.e. $\Ric_F \gg 1$, see equation \ref{eqn:nondimparams-for}).
A quantitative analysis of the simulation data for these cases showed that results in the weakly rotating limit (i.e. when the predicted Rossby number of the large-scale flow $\Ros_F$ defined in equation (\ref{eqn:nondimparams-for}) is greater or equal to 5) are nearly identical  to those of non-rotating simulations by \cite{garaud_turbulent_2016}. In particular, we found that Zahn's nonlinear instability criterion \citep{zahn_rotational_1974}, namely $J\Pra<(J\Pra)_c \simeq 0.007$, still holds and that the model of \cite{garaud_turbulent_2017} for mixing by diffusive turbulent shear flows (corrected for a minor error, see equation \ref{eq:nuturbgagnier}), correctly predicts the turbulent viscosity measured in the DNS when the shear instability is present. By contrast, the model of \cite{barker_angular_2019} for the GSF instability would largely under-predict momentum transport for the same simulations. And yet, we also discovered that the shear instability is not necessarily always excited when $J\Pra<(J\Pra)_c$. This is a fundamental difference between linear and nonlinear instabilities that should always be kept in mind -- nonlinear instabilities require a finite amplitude ``primer'' of the right form to develop, otherwise the instability does not take place. When that is the case, the GSF instability dominates instead and the  \cite{barker_angular_2019} model correctly predicts the turbulent viscosity. 

 At larger rotation rates (i.e. for a predicted Rossby number $\Ros_F$ of order unity or less), the GSF instability is increasingly dominant, and the turbulent viscosity measured in the DNS is consistent with the  \cite{barker_angular_2019} model. At the same time, the shear instability does not appear to be active, but even if it were, the predicted turbulent viscosity from the shear model of \cite{garaud_turbulent_2017} would be much smaller than that predicted by the GSF model of \cite{barker_angular_2019}, and would therefore be irrelevant.

These results pose an important question for stellar astrophysics. Indeed, stellar evolution codes usually compute turbulent mixing coefficients based on the local properties of the star (rotation rate, shear, stratification, etc.) -- where local here means local both in space and time. When a single instability is present, the physics that need to be included in the construction of that turbulent mixing coefficient are usually clear (whether they are all taken into account is another matter of course). However, when multiple instabilities are present at the same time, as is the case here, we see that the answer is significantly less obvious. One could simply focus on the instability that has the largest linear growth rate and ignore the others. However, this procedure would ignore all subcritical instabilities (whose linear growth rate is zero or negative), and could potentially underestimate the true turbulent viscosity by orders of magnitude, as seen in Section \ref{sec:barkercomp}. Ignoring potential interactions between the various instabilities, one could alternatively compute the turbulent viscosity associated with each of them individually (taking into account, this time, subcritical instabilities), and either add them all as is done in MESA for instance \citep{Paxton_MESAcode_2011}, or take their maximum value\footnote{This, in our opinion, makes somewhat more physical sense, as it assumes that one instability ends up dominating all the other ones. In practice the difference is not too significant.}.  
This approach is somewhat better supported by our data, but ignores the fact that subcritical instabilities are not always necessarily excited -- see the discussion in Sections \ref{sec:outliers} and \ref{sec:priming}. This leads to the fundamental question of {\it how can we predict whether the nonlinear shear instability is excited or not?} 

On that particular matter, an interesting outcome of our investigation is the discovery that the GSF instability can serve as a primer for the shear instability, provided the former does not saturate into a purely two-dimensional flow (see Section \ref{sec:priming}). In other words, the turbulence associated with the nonlinear saturation of the GSF instability (which draws its energy from the unstable angular momentum gradient) can sometimes also tap into the shear itself, and further drive diffusive shear instabilities. To do so, the GSF flow must be fully three-dimensional, which is usually the case unless $R_0$ defined in equation (\ref{eq:R0critGSF}) is very close to the marginal stability threshold for the GSF (i.e. unless $R_0 \rightarrow \Pra^{-1}$, see equation \ref{eq:instabscrit}).

With this in mind, we now propose the following algorithm to model the turbulent viscosity when both GSF and shear instabilities are potentially present {\it at the same time}\footnote{Regions of positive angular momentum gradient, where the GSF is not excited, must be treated differently.}: (1) Compute $R_0$ using (\ref{eq:R0critGSF}), and if $1<R_0 < \Pra^{-1}$ compute the associated turbulent viscosity for the GSF instability $\nu_{\rm GSF}$ (using equation \ref{eq:nuturbbarkerdim}). (2) Compute $J\Pra$, and if $J\Pra<0.007$ compute the turbulent viscosity associated with diffusive shear instabilities $\nu_{\rm shear}$ (using equation \ref{eq:nuturbgagnier}). 
(3) If $R_0$ is close to $\Pra^{-1}$, the flow is likely two-dimensional and will remain in a GSF state, hence use the computed value of the turbulent viscosity $\nu_{\rm GSF}$ (based on the discussion in Section \ref{sec:outliers}). (4) If $R_0$ is substantially lower than $\Pra^{-1}$, so the GSF instability is three-dimensional, then let $\nu_{\rm turb} = {\rm max} (\nu_{\rm shear},\nu_{\rm GSF})$. This also identifies the dominant instability in the flow. 

In this algorithm, the only missing ingredient is the threshold (in terms of $R_0$) beyond which the GSF instability remains two-dimensional. Finding this threshold will require a better understanding of the nonlinear saturation mechanism for the GSF instability at large $R_0$. From the numerical experiments shown in this paper at $\Pra=0.001$, we found that the GSF flow stays two-dimensional when $R_0$ exceeds about $500 = 0.5\Pra^{-1}$ (see in particular Figure \ref{fig:plotmodels}b). Whether a similar rule applies when $\Pra$ is much smaller remains to be determined.

Of course, much more work remains to be done before one can gain a complete understanding of shear instabilities and GSF instabilities in stars. In particular, the present study was limited to the equatorial region of a star, and the dynamics away from the equator are known to be substantially different, both for the GSF instability \citep{Knobloch_stability_1982,barker_angular_2020}, and for the shear instability \citep{cope2020dynamics,Garaud2020}. In the former case, \citet{barker_angular_2020} found that the GSF instability criterion is relaxed compared with (\ref{eqn:gsfdim}), and that angular momentum layering can lead to a vast increase in the turbulent transport coefficient. In the latter case, note that a star can undergo horizontal shear off equator, and \citet{cope2020dynamics} and \citet{Garaud2020} found that the horizontal shear instability criterion is much less stringent than that of the vertical shear instability, and that transport can be quite efficient. As such, it is quite likely that most our results do not apply at higher latitudes. 
In addition, most stars are magnetized, and the role of magnetic fields in suppressing or enhancing transport is highly non-trivial \citep[see, e.g.][]{tobias_betaplane_2007,Harrington_Garaud_2019,Chen_PVmixing_2020}. Nevertheless, our study has shed light on what might happen when these two very different instabilities coexist, demonstrating that it can lead to a variety of interesting and, in the case of the relaxation oscillations, potentially observable phenomena.




\section*{Acknowledgements}

The simulations were performed using the PADDI code kindly provided by Stephan Stellmach, on the XSEDE supercomputing facility Comet. This work is funded by NSF-AAG 1814327. We thank the referee, Adrian Barker, for insightful questions and comments that greatly helped improve this manuscript.

\section*{Data Availability}

The data published in this paper (table, figures) are available upon request to the first author. 



\bibliographystyle{mnras}
\bibliography{mnras_template.bib} 








\bsp	
\label{lastpage}
\end{document}